\documentclass[fleqn,usenatbib]{mnras}
\usepackage{newtxtext,newtxmath}
\usepackage[T1]{fontenc}

\DeclareRobustCommand{\VAN}[3]{#2}
\let\VANthebibliography\thebibliography
\def\thebibliography{\DeclareRobustCommand{\VAN}[3]{##3}\VANthebibliography}
\usepackage{graphicx}
\usepackage{amsmath}

\usepackage{multirow}
\usepackage{tabularx}

\newcolumntype{R}{>{\raggedleft\arraybackslash}X}
\usepackage{CJKutf8}
\newcommand{\orcid}[2]{\href{http://orcid.org/#2}{#1}}
\newcommand\wavemin{4427}
\newcommand\wavemax{4548}

\title[Stellar Populations over 6 Gyr]{The LEGA-C and SAMI Galaxy Surveys: Quiescent Stellar Populations and the Mass--Size Plane across 6\,Gyr}

\author[T. M. Barone et al.]{\orcid{Tania M. Barone}{0000-0002-2784-564X}$^{1,2,3,4}$\thanks{E-mail: tbarone@swin.edu.au},
\orcid{Francesco D'Eugenio}{0000-0003-2388-8172}$^{5,6}$,
\orcid{Nicholas Scott}{0000-0001-8495-8547}$^{2,4}$,
\orcid{Matthew Colless}{0000-0001-9552-8075 }$^{1,4}$,
\orcid{Sam P. Vaughan}{0000-0003-2265-7727}$^{2,3,4}$,
\newauthor
\orcid{Arjen van der Wel}{0000-0002-5027-0135}$^{5}$,
\orcid{Amelia Fraser-McKelvie}{0000-0001-9557-5648}$^{4,7}$,
\orcid{Anna de Graaff}{0000-0002-2380-9801}$^{8}$,
\orcid{Jesse van de Sande}{0000-0003-2552-0021}$^{2,4}$,
\orcid{Po-Feng Wu~\begin{CJK*}{UTF8}{bkai}(吳柏鋒)\end{CJK*}}{0000-0002-9665-0440}$^{9}$,
\newauthor
\orcid{Rachel Bezanson}{0000-0001-5063-8254}$^{10}$,
\orcid{Sarah Brough}{0000-0002-9796-1363}$^{4,11}$,
\orcid{Eric Bell}{0000-0002-5564-9873}$^{12}$,
\orcid{Scott M. Croom}{0000-0003-2880-9197}$^{2,4}$,
\orcid{Luca Cortese}{0000-0002-7422-9823}$^{4,7}$,
\orcid{Simon Driver}{0000-0001-9491-7327}$^{7}$,
\newauthor
\orcid{Anna R. Gallazzi}{0000-0002-9656-1800}$^{13}$,
\orcid{Adam Muzzin}{0000-0002-9330-9108}$^{14}$,
\orcid{David Sobral}{0000-0001-8823-4845}$^{15}$,
\orcid{Joss Bland-Hawthorn}{0000-0001-7516-4016}$^{2,4}$,
\orcid{Julia J. Bryant}{0000-0003-1627-9301}$^{2,4,16}$,
\newauthor
Michael Goodwin$^{16}$,
\orcid{Jon S. Lawrence}{0000-0002-6998-6993}$^{17}$,
\orcid{Nuria P. F. Lorente}{0000-0003-0450-4807 }$^{17}$,
\orcid{Matt S. Owers}{0000-0002-2879-1663}$^{18,19}$
\\
$^{1}$Research School of Astronomy and Astrophysics, The Australian National University, Canberra, ACT 2611, Australia\\
$^{2}$Sydney Institute for Astronomy, School of Physics, The University of Sydney, NSW, 2006, Australia\\
$^{3}$Centre for Astrophysics and Supercomputing, Swinburne University of Technology, P.O. Box 218, Hawthorn, VIC 3122, Australia\\
$^{4}$ARC Centre of Excellence for All Sky Astrophysics in 3 Dimensions (ASTRO 3D), Australia\\
$^{5}$Sterrenkundig Observatorium, Universiteit Gent, Krijgslaan 281 S9, B-9000 Gent, Belgium\\
$^{6}$Cavendish Laboratory and Kavli Institute for Cosmology, University of Cambridge, Madingley Rise, Cambridge, CB3 0HA, United Kingdom\\
$^{7}$International Centre for Radio Astronomy Research, The University of Western Australia, 35 Stirling Hwy, 6009 Crawley, WA, Australia\\
$^{8}$Leiden Observatory, Leiden University, P.O.Box 9513, NL-2300 AA Leiden, The Netherlands\\
$^{9}$Academia Sinica Institute of Astronomy and Astrophysics, No. 1, Section 4, Roosevelt Road, Taipei 10617, Taiwan\\
$^{10}$Department of Physics and Astronomy and PITT PACC, University of Pittsburgh, Pittsburgh, PA 15260, USA\\
$^{11}$School of Physics, University of New South Wales, NSW 2052, Australia\\
$^{12}$Department of Astronomy, University of Michigan, 1085 S. University Ave., Ann Arbor, MI, 48109, USA\\
$^{13}$INAF - Osservatorio Astrofisico di Arcetri, Largo Enrico Fermi 5, I-50125 Firenze, Italy\\
$^{14}$Department of Physics and Astronomy, York University, 4700 Keele St., Toronto, Ontario, M3J 1P3, Canada\\
$^{15}$Department of Physics, Lancaster University, Lancaster LA1 4YB, UK\\
$^{16}$Australian Astronomical Optics - USyd, University of Sydney, NSW 2006, Australia\\
$^{17}$Australian Astronomical Optics - Maquarie, Macquarie University, NSW 2109, Australia\\
$^{18}$Department of Physics and Astronomy, Macquarie University, NSW 2109, Australia\\
$^{19}$Astronomy, Astrophysics and Astrophotonics Research Centre, Macquarie University, Sydney, NSW 2109, Australia\\}

\date{Accepted XXX. Received YYY; in original form ZZZ}
\pubyear{2021}

\begin{document}

\defcitealias{Barone2018}{B18}
\defcitealias{Barone2020}{B20}

\label{firstpage}
\pagerange{\pageref{firstpage}--\pageref{lastpage}}
\maketitle

\begin{abstract}

We investigate changes in stellar population age and metallicity ([Z/H]) scaling relations for quiescent galaxies from intermediate redshift ($0.60\leq~z\leq0.76$) using the LEGA-C Survey to low redshift ($0.014\leq~z\leq0.10$) using the SAMI Galaxy Survey. Specifically, we study how the spatially-integrated global age and metallicity of individual quiescent galaxies vary in the mass--size plane, using the stellar mass $M_*$ and a dynamical mass proxy derived from the virial theorem $M_D\propto\sigma^2~R_\mathrm{e}$. We find that, similarly to at low-redshift, the metallicity of quiescent galaxies at $0.60\leq~z\leq 0.76$ closely correlates with $M/R_\mathrm{e}$ (a proxy for the gravitational potential or escape velocity), in that galaxies with deeper potential wells are more metal-rich. This supports the hypothesis that the relation arises due to the gravitational potential regulating the retention of metals by determining the escape velocity for metal-rich stellar and supernova ejecta to escape the system and avoid being recycled into later stellar generations. Conversely, we find no correlation between age and surface density ($M/R_\mathrm{e}^2$) at $0.60\leq~z\leq 0.76$, despite this relation being strong at low-redshift. We consider this change in the age--$M/R_\mathrm{e}^2$ relation in the context of the redshift evolution of the star-forming and quiescent mass--size relations, and find our results are consistent with galaxies forming more compactly at higher redshifts and remaining compact throughout their evolution. Furthermore, galaxies appear to quench at a characteristic surface density that decreases with decreasing redshift. The $z\sim~0$ age--$M/R_\mathrm{e}^2$ relation is therefore a result of building up the quiescent and star-forming populations with galaxies that formed at a range of redshifts and therefore a range of surface densities.

\end{abstract}


\section{Introduction}

A challenge in the field of galaxy evolution is understanding the influence of redshift-dependent Universe conditions; i.e., how does the redshift at which a galaxy formed affect its evolutionary path (its total mass-assembly and star-formation history)? Galaxies observed at $z \sim 0.76$ (nearly 7\,Gyr lookback time) are remarkably different to those observed in the present-day Universe \citep[e.g.][]{Stott2016,vanderWel2016,Wisnioski2019}. Compared to nearby galaxies at fixed stellar mass, at intermediate redshifts ($z \sim 0.5$--3) galaxies are more compact \citep{Ferguson2004,Trujillo2007,Buitrago2008,Williams2010,vanderWel2014,Mowla2019}, more highly star-forming \citep[with a peak in cosmic star formation density at $z \approx 2$;][]{Lilly1996,Madau1998,Daddi2007,Madau_Dickinson2014}, have dynamically hotter turbulent star-forming gas \citep{ForsterSchreiber2006,Weiner2006,Law2007,ForsterSchreiber2009,Law2009,Wright2009,Wisnioski2011,Epinat2012,Swinbank2012,Kassin2012,Wisnioski2015}, and higher molecular gas fractions \citep{Tacconi2010,Daddi2010,Tacconi2013,Morokuma-Matsui2015}, indicating significant evolution over this time period. Crucially, galaxies today will not follow the same evolutionary path over the next 6\,Gyr as similarly-massive galaxies that formed 6\,Gyr ago \citep[e.g.][]{Barro2013,Abramson2016}, because the conditions that influence a galaxy's evolutionary path (e.g. the availability of pristine gas, the scale of galaxy clustering) change as the Universe evolves.

The size difference between star-forming and quiescent galaxies is one piece of evidence for this change in evolutionary path. At fixed mass, star-forming galaxies are larger than their quiescent counterparts \citep{Kriek2009,Williams2010,Wuyts2011,vanderWel2014,Whitaker2017}. While passive disk fading after ceasing star formation does lead to a decrease in effective radius, this process alone is insufficient to explain the size  difference between the $z=0$ star-forming and quiescent populations \citep{Croom2021_disk_fading}. Therefore, this difference in average size indicates that the star-forming progenitors of present day quiescent galaxies were different to the $z\sim0$ population of star-forming galaxies, which will instead evolve into extended quiescent galaxies \citep{Barro2013}.

Understanding the redshift dependence of galaxy evolution requires disentangling the evolution of an individual galaxy within a population from the evolution in the average properties of the population as a whole (due to the continual addition of newly formed galaxies with different properties to the extant population). A key method of studying how both individual and populations of galaxies evolve is by analysing their growth in mass and size. An individual star-forming galaxy is expected to evolve along the relation of star-forming galaxies in the mass--size plane \citep{Lilly1998,Ravindranath2004,Trujillo2006_apj,Pezzulli2015,vanDokkum2015}, while the average size of the population as a whole increases with decreasing redshift due to processes such as mergers \citep{Hopkins2009,Naab2009} and "puffing up" due to strong AGN feedback \citep{Fan2008,Fan2010}, but also new galaxies forming with larger radii \citep{vanDokkum_Franx2001,Carollo2013}. In other words, star-forming galaxies at both low and intermediate redshifts follow parallel tracks in the mass--size plane \citep{Speagle2014,vanDokkum2015}, with their starting location in the plane depending on redshift. Therefore, when analysing star-forming galaxies in the mass--size plane at $z\sim0$, we are observing the combined effect of both the evolution of individual galaxies throughout their lifetimes and the evolution of the population due to the addition of new members and loss of old members (as they quench). We want to disentangle these two effects to understand how the redshift range over which a galaxy formed and evolved influences its evolutionary path. We therefore need to understand how the processes influencing a galaxy may be regulated by the broader conditions of the Universe and how these conditions change with redshift. An important tool to measure the impact of various processes is the analysis of scaling relations, which quantify the link between different galaxy parameters to determine their dependence. Specifically, scaling relations between stellar population parameters and galaxy structure allow us to quantify how processes involved in star-formation and stellar mass assembly interrelate with processes dominating structural and dynamical changes.

Recent studies analysing various stellar population scaling relations have demonstrated a clear dependence of  stellar population and star formation history on galaxy size ($R_{\mathrm{e}})$, at both low and intermediate ($z\sim 0 - 3$) redshifts \citep[e.g.,][]{Bell2000,Bell_deJong2000,Kauffmann2003_II,vanderWel2009}. \cite{Franx2008} investigated the dependence of $u-g$ colour (interpreted as a proxy for star-formation history) on stellar mass ($M_*$) and $R_{\mathrm{e}}$ for a sample spanning $0<z<3$. At all redshifts, they showed that $M_*$ alone is not a good predictor of colour (and thus star-formation history), and that the correlations between $u-g$ colour and $M_*/R_\mathrm{e}^2$ or $M_* / R_\mathrm{e}$ have less scatter than the relations with $M_*$. \cite{Wake2012} extended this work at low redshift ($z<0.11$), showing that $u-r$ colour correlates more strongly with velocity dispersion ($\sigma$) than $M_*$, $M_* / R_\mathrm{e}^2$ or S\'{e}rsic index \citep{Sersic1968}. More recently, \cite{Diaz-Garcia2019} used spectral energy distribution fits to optical and near-infrared photometry from ALHAMBRA \citep[Advanced Large, Homogeneous Area Medium Band Redshift Astronomical Survey;][]{Moles2008}, to study how the stellar population properties of quiescent galaxies vary in the mass--size plane up to redshift $z\sim 1$. They found that stellar population properties show a dependence on galaxy mass and size since $z\sim 1$. Furthermore, low-redshift studies by \cite{McDermid2015}, \cite{Scott2017} and \cite{Li2018} using spectroscopically-derived stellar population parameters showed that much of the scatter in the age--$M_*$ and metallicity--$M_*$ relations is due to residual trends with galaxy size, in that smaller galaxies at fixed mass are older and more metal rich. This difference in the resulting mean stellar population parameters can be traced back to differences in the star formation histories using simulations. \cite{Gupta2021} used the IllustrisTNG simulations \citep{Pillepich2018} to track the evolution of extended massive galaxies and found that, compared to their normal-sized counterparts, extended galaxies quench later despite having similar star-formation rates and stellar masses when selected at $z\sim2$. Furthermore, in addition to a dependence of global stellar population on galaxy structure, studies have also found a dependence \textit{within} galaxies between local stellar population properties and local dynamical and structural parameters \citep[e.g.][]{Gonzalez_Delgado2014_II,Gonzalez_Delgado2015,MollerChristensen2019,Zibetti2020}, which suggests the global relations arise from local scales \citep{Scott2009}.

\citet[][hereafter \citetalias{Barone2018} and \citetalias{Barone2020} respectively]{Barone2018,Barone2020} built on these earlier studies by quantifying this observed dependence on size, and showing how global age and metallicity ([Z/H]) correlate with the galaxy structure: $M_*/R_\mathrm{e}$ (a proxy for the gravitational potential, $\Phi$, or escape velocity\footnote{As shown by \cite{Scott2009} and \cite{Cappellari2013a}, the difference between the escape velocities for the stellar body and the dark matter halo is not large ($\sim$0.1~dex), hence we can adopt $M_*/R_\mathrm{e}$ as a qualitative but reliable proxy for the potential well depth and escape velocity.)}) and $M_*/R_\mathrm{e}^2$ (stellar mass surface density, $\Sigma$) for low-redshift early-type and star-forming galaxies. Specifically, these studies quantified and compared the intrinsic scatter within each relation and any residual trend with galaxy size. Despite using different samples, methods and models, the two studies found the same results in both early-type and star-forming galaxies: the correlations that are the tightest and have the least residual trend with galaxy size are the age--$\Sigma$ and [Z/H]--$M_*/R_\mathrm{e}$ relations. Additionally, \cite{DEugenio2018} showed that gas-phase metallicity in star-forming galaxies at low redshift is also more tightly correlated with $M_*/R_\mathrm{e}$ than either $M_*$ or $\Sigma$.

Based on these results, \citetalias{Barone2018}, \citetalias{Barone2020} and \cite{DEugenio2018} proposed and discussed various mechanisms that could lead to the [Z/H]--$\Phi$ and age--$\Sigma$ relations. These studies concluded that the [Z/H]--$\Phi$ relation is driven by galaxies with low gravitational potentials losing more of their metals because the escape velocity required for metal-rich gas to be expelled by supernova feedback is directly proportional to the depth of the gravitational potential. Given this assumption, there should also exist a correlation between metallicity and gravitational potential at intermediate redshifts, although the slope and scatter of the relation may vary due to changes in the strength of star formation feedback and outflows. As for the age--$\Sigma$ relation, \citetalias{Barone2020} proposed it results from compact galaxies having formed earlier than their diffuse counterparts, and so the mechanism(s) responsible for determining the size and mass of a galaxy at fixed age depend on redshift. This hypothesis is in agreement with the results of \cite{Diaz-Garcia2019}, who found that the formation epoch of quiescent galaxies since $z \sim 1$ shows a strong dependence on size at fixed mass. The age--$\Sigma$ relation may therefore be less pronounced at intermediate redshifts, as less time has passed for the relation to build up. The aim of this paper is to test these hypotheses, by determining the redshift dependence of the age--$\Sigma$ and [Z/H]--$\Phi$ scaling relations.

In this paper we build on the results of stellar populations scaling relations at low redshift by analysing if (and how) the relations change across a lookback time of 6\,Gyr ($0.014 \leq z \leq 0.76$), to test the hypotheses proposed for the age--$\Sigma$ and [Z/H]--$\Phi$ scaling relations. Specifically, we study how the spatially-integrated average (global) age and metallicity of individual quiescent galaxies vary in the mass--size plane, using both the stellar mass $M_*$ and a proxy for the dynamical mass derived from the virial theorem ($M_D \propto \sigma^2 R_\mathrm{e}$). Furthermore, we look at how the age--$\Sigma$ and [Z/H]--$\Phi$ relations found at low redshift appear at $0.60 \leq z \leq 0.76$. By quantifying the significance of these scaling relations at low and intermediate redshift we aim to understand their origins and, in the process, begin to discern the redshift dependence of stellar population evolution over the past 6\,Gyr. 

The paper is arranged as follows. In Section~\ref{sec:sample_selection} we describe the $0.60 \leq z \leq 0.68$ and $0.68 < z \leq 0.76$ samples from the LEGA-C survey and the $0.014 \leq z \leq 0.10$ comparison sample from the SAMI survey. Section \ref{sec:methodology} describes the full spectral fitting method used to obtain the metallicity and age measurements, as well as our analysis methods. We present the [Z/H] and age results in Section~\ref{sec:results}, followed by a discussion in Section~\ref{sec:discussion}. In discussing the age results we link the age--$\Sigma$ relation to the distribution of the quiescent and star-forming populations across redshift in the mass--size plane in Section~\ref{Population evolution in the Mass--Size Plane}. Lastly we provide a summary of our conclusions in Section~\ref{sec:Summary and Conclusions}. We assume a flat $\Lambda$ cold dark matter ($\Lambda$CDM) Universe with $\Omega_{\Lambda}=0.7$, $\Omega_M = 0.3$ and $H_0 = 70$\,km\,s\textsuperscript{$-1$}\,Mpc\textsuperscript{$-1$}, and a \cite{Chabrier2003} initial mass function.

\section{Data}\label{sec:sample_selection}

We describe the LEGA-C and SAMI surveys in Sections~\ref{subsec:legac_survey_sample} and~\ref{subsec:sami_survey_sample} respectively, followed by describing auxiliary parameter measurements in Section~\ref{subsec:measurements of galaxy properties}. The adopted quiescent versus star-forming galaxy selection criterion is detailed in Section~\ref{sec:quiescent_galaxy_selection}.

\subsection{The LEGA-C Survey}\label{subsec:legac_survey_sample}

The Large Early Galaxy Astrophysics Census \citep[LEGA-C; ][]{vanderWel2016} is a slit spectroscopic survey of galaxies at intermediate redshifts in the COSMOS field \citep{Scoville2007}, using the VIsible Multi-Object Spectrograph \citep[VIMOS; ][]{LeFevre2003} on the Very Large Telescope. LEGA-C targets were selected from the UltraVISTA catalogue \citep{Muzzin2013_catalogue} based on a redshift-dependent apparent $K_S$ magnitude limit: $K_{s} < 20.7 - 7.5 \log((1 + z)/1.8)$ \citep{vanderWel2016}, which has the advantage of closely resembling a selection on stellar mass while remaining model independent \citep[see Appendix A of][]{vanderWel2021}. The survey comprises 4209 galaxies, 3472 of which have spectroscopic redshift measurements within $0.60 \leq z \leq 1.0$. Targets were observed for $\sim$20 hours to reach an approximate S/N per {\AA} $\approx 20$ in the continuum. We use integrated spectra summed along the entire slit, with an effective spectral resolution of R$\sim$3500 \citep{Straatman2018} and observed wavelength range $\sim$6300{\AA}–8800{\AA}. We use spectra from the third data release \citep{vanderWel2021}\footnote{The data can be accessed from \href{http://archive.eso.org/cms/eso-archive-news/Third-and-final-release-of-the-Large-Early-Galaxy-Census-LEGA-C-Spectroscopic-Public-Survey-published.html}{http://archive.eso.org/cms/eso-archive-news/Third-and-final-release-of-the-Large-Early-Galaxy-Census-LEGA-C-Spectroscopic-Public-Survey-published.html}}; see \cite{vanderWel2016} and \cite{Straatman2018} for earlier data releases.

While the LEGA-C sample spans the redshift range $0.60 \leq z \leq 1.0$, above $z \sim 0.8$ the survey selection criteria and the S/N requirement for stellar population analyses (median S/N per {\AA} $\geq 10$ in the rest-wavelength region \wavemin--\wavemax~{\AA}; see Section \ref{sec:survey_comparison}) limit the sample to only the brightest (most massive) targets. We therefore restrict our analysis to the redshift range $0.60 \leq  z \leq 0.76$. We then estimate the size evolution with redshift for this redshift range using the relations from \cite{vanderWel2014} and \cite{Mowla2019} for high ($\log M_* / M_{\odot} > 11.3$) and intermediate $\log M_* / M_{\odot} \sim 10.75$) mass galaxies respectively. To ensure the size evolution is less than 0.05~dex (the uncertainty on the size measurements), we split the sample into two redshift bins: $0.60 \leq z \leq 0.68$ and $0.68 < z \leq 0.76$. Lastly, given the LEGA-C sample at z$\sim 0.7$ is representative down to a stellar mass limit of $\log M_* \geq 10.48$ \citep{vanderWel2021}, we restrict our analysis to LEGA-C galaxies with $10.48 \leq \log M_* \leq 11.5$. This leaves final samples of 412 galaxy (219 quiescent, 193 star-forming) for $0.60 \leq z \leq 0.68$, and 513 (273 quiescent, 240 star-forming) for $0.68 < z \leq 0.76$.

\subsection{The SAMI Galaxy Survey}\label{subsec:sami_survey_sample}

The Sydney-AAO Multi-object Integral-field \citep[SAMI;][]{Bryant2015,Croom2021_SAMIDR3} Galaxy Survey is a low-redshift integral-field survey of 3068 unique galaxies observed using the SAMI instrument \citep{Croom2012} connected to the AAOmega spectrograph \citep{Sharp2006, Sharp2015} on the Anglo-Australian Telescope. The sample spans the redshift range $0.004 \leq z \leq 0.11$. Galaxies were observed through fused-fibre hexabundles \citep{Bland-Hawthorn2011,Bryant2014}, each comprising 61 tightly-packed fibres that form an approximately circular grid with a diameter of 15". Observations were typically for a total of 3.5~hours comprised of 7~dithers of 30~minutes each. Targets were selected based on stellar mass cuts in narrow redshift bins from three equatorial regions covered by the volume-limited Galaxy And Mass Assembly \citep[GAMA; ][]{Driver2011} survey and 8 cluster regions; see \cite{Bryant2015} and \cite{Owers2017} for a full description of the sample selection for the GAMA and cluster regions respectively. For this analysis we exclude targets from the cluster regions to avoid over-representing galaxies from dense environments \citep[e.g. as discussed in][]{vandeSande2021}, leaving the SAMI-GAMA sample which is highly complete ($\sim90\%$) for $\log M_* \geq 10$ \citep{Bryant2015,vandeSande2021, Croom2021_SAMIDR3}. SAMI spectra have two components, a `blue' component (3700--5700\,{\AA}) at a resolution of R=1800, and a `red' component (6300--7400\,{\AA}) at R=4300. For the full-spectral fits the higher spectral resolution red component is degraded to match the resolution of the blue component. The whole spectrum is then fit simultaneously. We use spatially-integrated $1 R_\mathrm{e}$ spectra from the third data release \citep{Croom2021_SAMIDR3}\footnote{The data can be accessed at \href{https://docs.datacentral.org.au/sami/}{https://docs.datacentral.org.au/sami/}}; see \cite{Allen2015}, \cite{Green2018} and \cite{Scott2018} for earlier releases. 

As for the LEGA-C samples, we want to ensure the redshift evolution within our SAMI sample is less than the typical uncertainty on the size measurements (0.05~dex). We calculate the estimated size evolution within the SAMI redshift range using the same relations of \cite{vanderWel2014} and \cite{Mowla2019} used in Section \ref{subsec:legac_survey_sample} for the LEGA-C sample. Based on the estimated size evolution, we remove 22 galaxies at the ends of the redshift range. Our final SAMI sample comprises of 974 galaxies (524 quiescent, 450 star-forming) with $10.0 \leq \log M_* \leq 11.5$, median S/N per {\AA} $\geq 10$ in the rest wavelength region \wavemin--\wavemax, and in the redshift range $0.014 \leq z \leq 0.10$.

\subsection{Measurements of galaxy properties}\label{subsec:measurements of galaxy properties}

\begin{figure}
    \centering
    \includegraphics[width=\columnwidth]{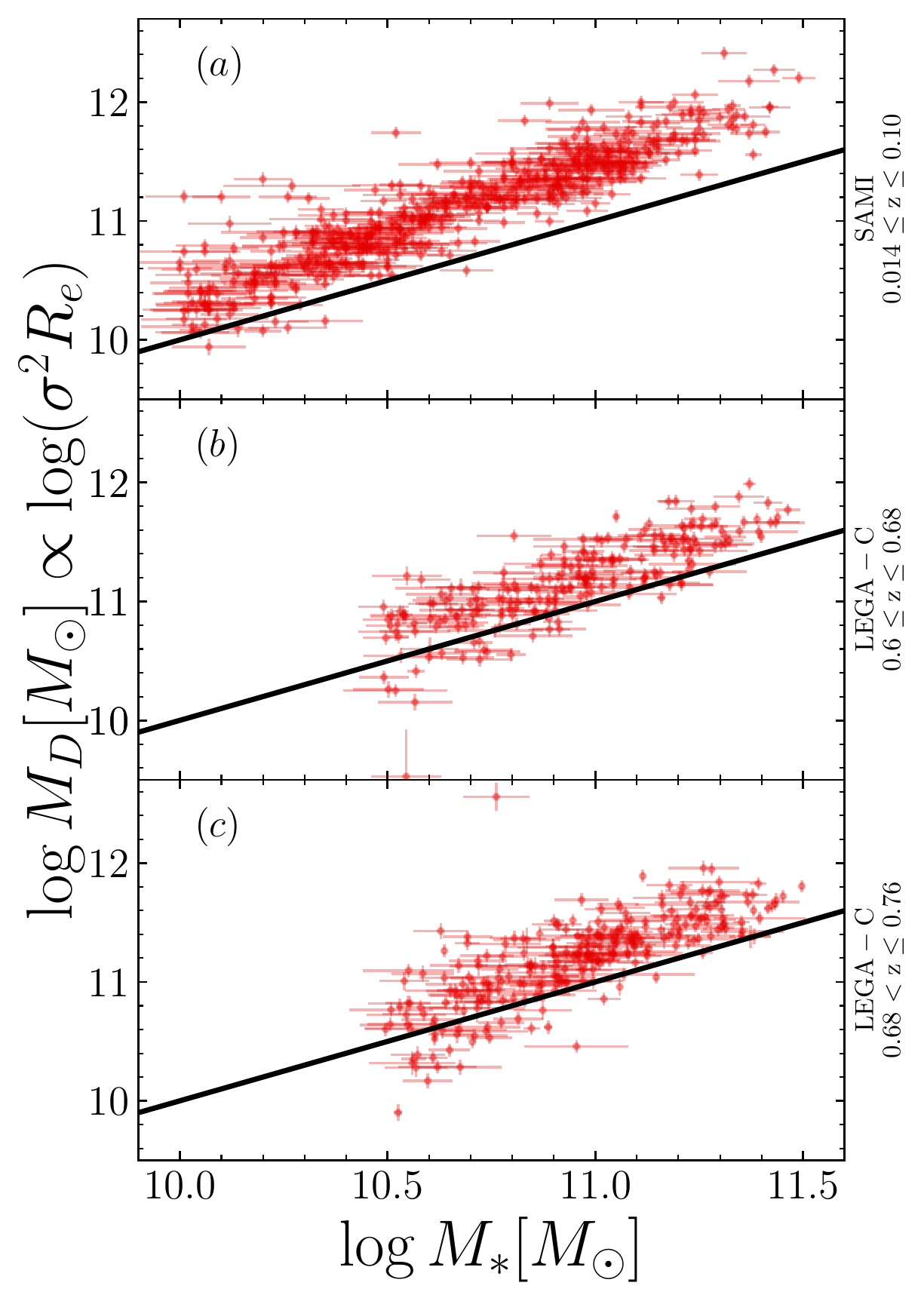}
    \caption{The stellar mass $M_*$ versus the virial proxy for the dynamical mass $M_D$ for the three redshift samples. The solid black line shows the 1-1 relation. Each point shows the $1\sigma$ uncertainty on $\log M_*$ and $\log M_D$. There is good agreement between the two mass estimates, with the dynamical masses being 0.4 and 0.2~dex higher for the SAMI and LEGA-C samples respectively.}
    \label{fig:stellar_vs_dynamical_mass}
\end{figure}

Semi-major effective radii ($R_\mathrm{e}$) were measured from S\'ersic fits using the \textsc{galfit} software \citep{Peng2010_galfit} by \cite{vanderWel2021} for the LEGA-C sample and \cite{Hill2011} for the SAMI sample. For LEGA-C targets \cite{vanderWel2021} followed the procedure of \cite{vanderWel2012} on HST ACS F814W images from the COSMOS program \citep{Scoville2007}. For SAMI targets, \cite{Hill2011} followed the process of \cite{Kelvin2012}, using $r$-band photometry from SDSS DR7 \citep{Abazajian2009} reprocessed for GAMA. We note that colour gradients within galaxies affect the absolute measurement of $R_\mathrm{e}$, in that bluer bands lead to larger size measurements \citep[e.g.][]{Kelvin2012}. However, given our analysis focuses on the relative difference in size \textit{within} each redshift bin, any systematic offset between the SAMI and LEGA-C size measurements due to the difference in photometric band does not affect our results.

We use stellar masses ($M_*$) and star formation rates (SFR) measured by \cite{deGraaff2020,deGraaff2021} for the LEGA-C galaxies and \cite{Driver2018} for the SAMI galaxies. In both catalogues the measurements were derived using the \textsc{magphys} \citep[Multi-wavelength Analysis of Galaxy Physical Properties;][]{daCunha2008_MAGPHYS} spectral energy distribution fitting software, based on the \cite{BruzualCharlot2003} library of simple stellar population spectra, a \cite{Charlot_Fall2000} dust attenuation model, a \cite{Chabrier2003} initial mass function and an exponentially declining star formation history (however the LEGA-C star formation history also includes random star-formation bursts). While the two catalogues differ slightly in the photometric bands used for the fits due to differences in data availability, they both include photometry ranging from ultra-violet to the far-infrared. Specifically, \cite{deGraaff2021} used 
ugriz and BVYJHK\textsubscript{s} from UltraVISTA as well as Spitzer infrared and multiband (rest-frame mid ultra-violet to far-infrared),  while \cite{Driver2018} used FUV, NUV, ugriz, ZYJHK, W1234, PACS100/160 and SPIRE 250/350/500 (rest-frame far ultra-violet to far-infrared). See \cite{deGraaff2021} and \cite{Driver2018} for further details. The samples are complete in stellar mass within each redshift bin ($\log M_* \geq 10, 10.48, 10.48 M_{\odot}$ for $z_{\mathrm{SAMI}} \in [0.014, 0.10]$, $z_{\mathrm{LEGA-C}} \in [0.60, 0.68], [0.68, 0.76]$ respectively). Our results are qualitatively unchanged if we use the same stellar mass limit ($\log M_* \geq 10.48$) for all three redshift bins.

We choose to use an independent measure of the stellar mass rather than those derived from the full spectral fits due to the increased wavelength coverage of the MAGPHYS masses. Given the rest spectral range of the LEGA-C ($\sim$3700–5200 {\AA} for $z=0.67$) and SAMI spectra ($\sim$3500–5400 and 6000–7000 {\AA} for $z=0.05$), masses derived from the full spectral fits include predominantly blue light, so they will be inevitably biased due to not sampling the rest-frame infrared. MAGPHYS stellar masses are based on a broader wavelength range (rest-frame ultra-violet to far infrared), which contains more information about dust extinction and reaches redder wavelengths than the spectroscopy alone.

In addition to covering the GAMA regions from which the SAMI sample was selected, the \cite{Driver2018} catalogue also includes measurements for the G10-COSMOS region \citep{Davies2015_G10,Andrews2017_G10} which includes 301 LEGA-C galaxies. We use this overlap of 301 galaxies to check for systematic biases between the samples that might arise from the different photometric bands used in the SED fits. We find good agreement between the two catalogues for both the stellar masses and the star formation rates. The mean difference between the two catalogues is $0.08 \pm 0.15$~dex for $\log M_*$ and $0.07 \pm 0.40$~dex for $\log \mathrm{SFR}$, with the values from \cite{deGraaff2021} being slightly more massive and star-forming.

Lastly we define a simple proxy for the total (dynamical) galaxy mass $M_D$ based on the virial theorem:
\begin{equation}
    M_D \equiv k\sigma^2 R_\mathrm{e}/G    
\end{equation}

Where $\sigma$ is the aperture velocity dispersion (measured from the \textsc{pPXF} fits; see Section \ref{sec:stellar_pops_measurement}), $G$ is the gravitational constant ($4.3\times 10^{-6}$\,km\textsuperscript{2}\,kpc\,s\textsuperscript{$-2$}\,$M_{\odot}^{-1}$), and $k$ is a constant set to $5.0$ \citep{Cappellari2006}. Figure \ref{fig:stellar_vs_dynamical_mass} shows the stellar and dynamical mass estimates for the quiescent galaxies in the three redshift bins. In all three redshift samples there is good linear agreement between the stellar and dynamical estimates, with $M_D$ being systematically 0.4 and 0.2~dex more massive than $M_*$ for SAMI and LEGA-C galaxies respectively. We note that the higher $\log M_D - \log M_*$ for SAMI galaxies compared to LEGA-C is likely driven by the smaller physical coverage of the SAMI galaxies ($1 R_\mathrm{e}$) compared to LEGA-C ($\gtrapprox 1 R_\mathrm{e}$; see Section \ref{sec:survey_comparison}). However, given our analysis focuses on trends within redshift bins, systematic differences between bins does not influence our results. For a detailed discussion and analysis of dynamical versus stellar masses for LEGA-C galaxies, we refer the reader to \cite{deGraaff2021}.

\begin{figure}
    \centering
    \includegraphics[width=\columnwidth]{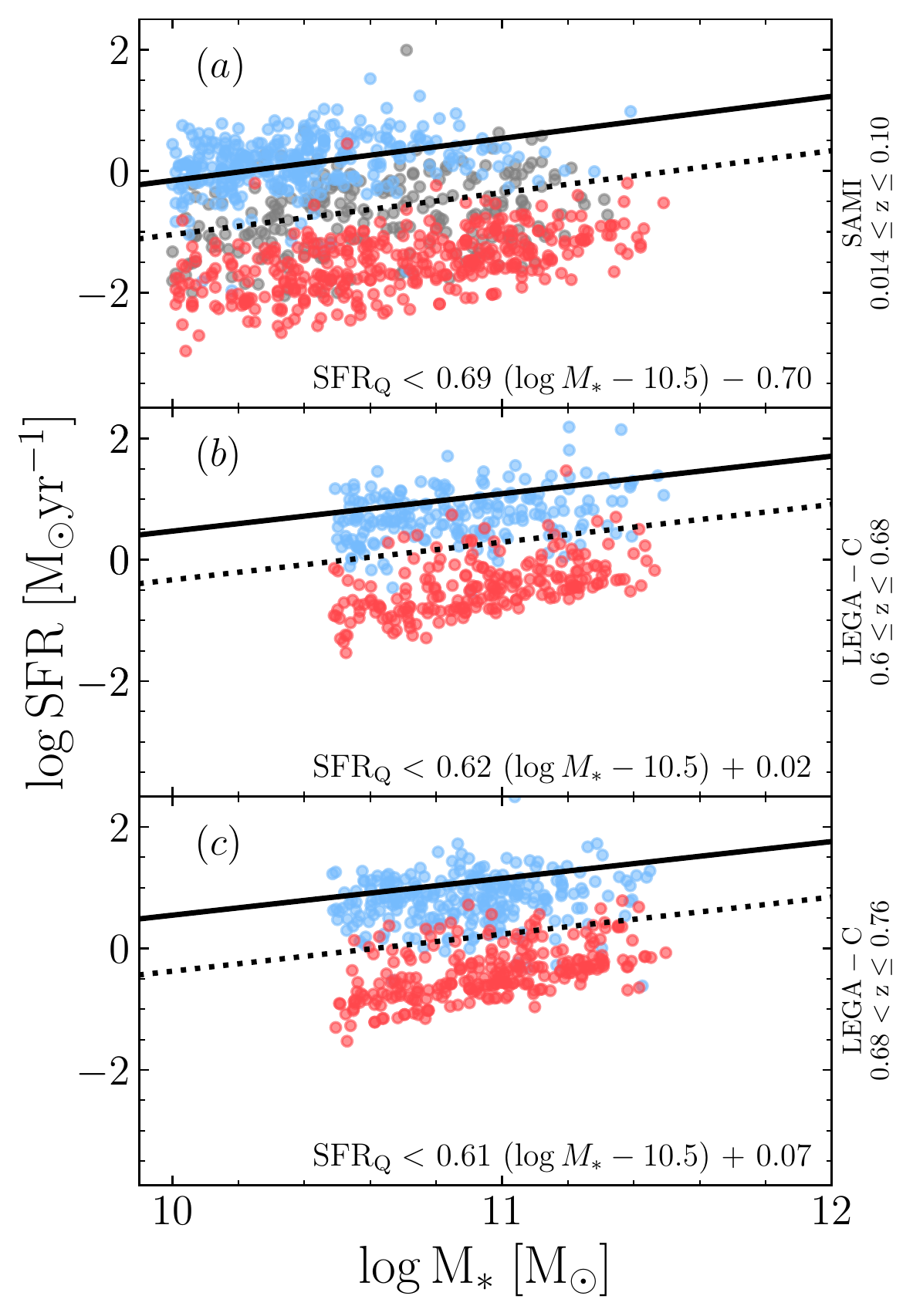}
    \caption[Quiescent galaxy sample selection using the star formation rate--stellar mass diagram for SAMI and LEGA-C galaxies]{Star formation rate versus stellar mass for the three redshift bins. The black solid lines are the redshift-dependent main sequence from \protect\citet[][see Equation~\ref{whitaker2012_MS}]{Whitaker2012}. The dotted black lines are the adopted boundary between quiescent and star-forming galaxies, defined as $ \mathrm{SFR}_Q < \mathrm{SFR}_{MS} - 2\mathrm{RMS}$ (bottom right of each panel), where RMS is the measured scatter about the main sequence based on an initial cut 1\,dex below the relation. The points are coloured using quiescent (red points), star-forming (blue points), and intermediate (grey points; panel a only) selection criteria from the literature (see Section~\ref{Comparison to other Quiescent classifications}); our adopted selection criteria are comparable to these alternatives.}
    \label{fig:sfr_quiescent_selection}
\end{figure}

\subsection{Quiescent Galaxy Selection}\label{sec:quiescent_galaxy_selection}

We separate quiescent and star-forming galaxies in both the LEGA-C and SAMI samples based on their distance from the redshift-dependent star-forming main sequence defined by \cite{Whitaker2012}:
\begin{equation}\label{whitaker2012_MS}
   \log \mathrm{SFR} = (0.70 - 0.13z)(\log M_* - 10.5) + 0.38 + 1.14z - 0.19z^2
\end{equation}

We show this relation in Figure~\ref{fig:sfr_quiescent_selection} (solid black line) using the central redshift value in each bin ($z = 0.057, 0.64, 0.72$). We do an initial cut $1$\,dex below the relation and measure the root-mean-square (RMS) scatter about the trend. We then define quiescent galaxies as those with $\mathrm{SFR}_Q < \mathrm{SFR}_{MS} - 2\mathrm{RMS}$ (dotted black line). The precise selection cut for each redshift bin is given in the corresponding panel of Figure~\ref{fig:sfr_quiescent_selection}.

Post-starburst galaxies are a subclass of quiescent galaxies that recently quenched after a burst of star-formation \citep{Dressler_Gunn1983,Balogh1999,Dressler1999}, and therefore contain a young average stellar population. As a result of their rapid quenching \citep{Wu2020}, post-starburst galaxies tend to be smaller than the rest of the quiescent population at fixed mass \citep{Whitaker2012,Yano2016,Almaini2017}. The young average ages and small sizes of post-starburst galaxies are contrary to the rest of the quiescent population in which young galaxies tend to be larger at fixed mass \citep{Wu2018}. The number of post-starburst galaxies in our three redshift samples (defined as quiescent galaxies with an equivalent width of $H\delta_A > 4${\AA}; e.g. \citealt{Wu2018,DEugenio2020}) is 2, 8, and 19 (0.4\%, 4\%, and 7\%) for $0.014 \leq z_\mathrm{SAMI} \leq 0.10$, $0.60 \leq z_\mathrm{LEGA-C} \leq 0.68$ and $0.68 < z_\mathrm{LEGA-C} \leq 0.76$ respectively. It can be argued that post-starburst galaxies should be removed from our sample, because their evolutionary path with a final starburst in the centre is different from the rest of the galaxy population 
\citep{Wu2018,Wu2020,DEugenio2020}. However, when we repeat our analysis after having removed these galaxies, we find the results qualitatively unchanged. Therefore we do not remove post-starburst galaxies from our quiescent samples.

\subsubsection{Comparison to other Quiescent classifications}\label{Comparison to other Quiescent classifications}
We use the star-forming main sequence as defined by \cite{Whitaker2012} because their analysis used the same cosmological parameters and IMF assumed in this work and covers the full redshift range of both the LEGA-C and SAMI samples. We note however that the \cite{Whitaker2012} results are based on linear fits in the $\log \mathrm{SFR}$--$\log M_*$ plane, but more recent works have shown that the star-forming main sequence turns over at high stellar masses \citep[e.g.][]{Leslie2020}. We therefore test our analysis using other quiescent selection criteria used in the literature for the two surveys.

In Figure \ref{fig:sfr_quiescent_selection} we compare our quiescent selection criterion to others used in the literature for the two surveys. The SAMI galaxies (panel~a) are coloured by the selection used in \cite{Croom2021_disk_fading}, which is based on the star formation rates from \cite{Medling2018} and distance from the \cite{Renzini_Peng2015} main sequence. The LEGA-C galaxies (panels~b and~c) are coloured by their U--V vs V--J diagram classification \citep{Labbe2005}, as defined by \cite{Muzzin2013_mass_function} and used in, e.g., \cite{Chauke2018}, \cite{Wu2018} and \cite{Straatman2018}. Figure~\ref{fig:sfr_quiescent_selection} shows that our quiescent selection based on distance from the \cite{Whitaker2012} main sequence is comparable to these alternative selection criteria used in the literature. We repeated our anlysis using these alternate quiescent selection criteria and find our results unchanged. Therefore, we use the quiescent criteria based on the \cite{Whitaker2012} main sequence and \textsc{magphys} derived SFR and $M_*$ for consistency across the two surveys.

\subsection{Survey Comparison}\label{sec:survey_comparison}

Galaxies have well established internal radial stellar population gradients that, when combined with a varying aperture size, can lead to spurious global trends. A key difference between the two surveys are the apertures used; the VIMOS instrument used by LEGA-C has slits 1"  wide and 8" long whereas the SAMI instrument uses fused-fibre hexabundles of 15" diameter, comprised of 61 individual fibres each 1.6" in diameter. For the SAMI targets we use spectra integrated within 1 effective radius ($R_\mathrm{e}$), and for the LEGA-C data we use spectra integrated along the entire slit. At $z=0.7$ 1" corresponds to a physical scale of $\sim$7.1~kpc which encompasses at least 50\% of the total flux, but depending on the galaxy's apparent size the spectrum will also contain flux from outer regions ($R > R_\mathrm{e}$). Therefore, despite the different apertures used in the two surveys, all galaxies are probed to at least $1 R_\mathrm{e}$, mitigating potential aperture bias in measurements between the two surveys. We emphasise, however, that we are not directly comparing stellar population parameters between the redshift bins, rather we are interested in how properties vary \textit{within} redshift bins. Therefore, any systematic bias in stellar population measurements arising from the differences between the two instruments do not influence our analysis.

\begin{figure}
    \centering
    \includegraphics[width=\columnwidth]{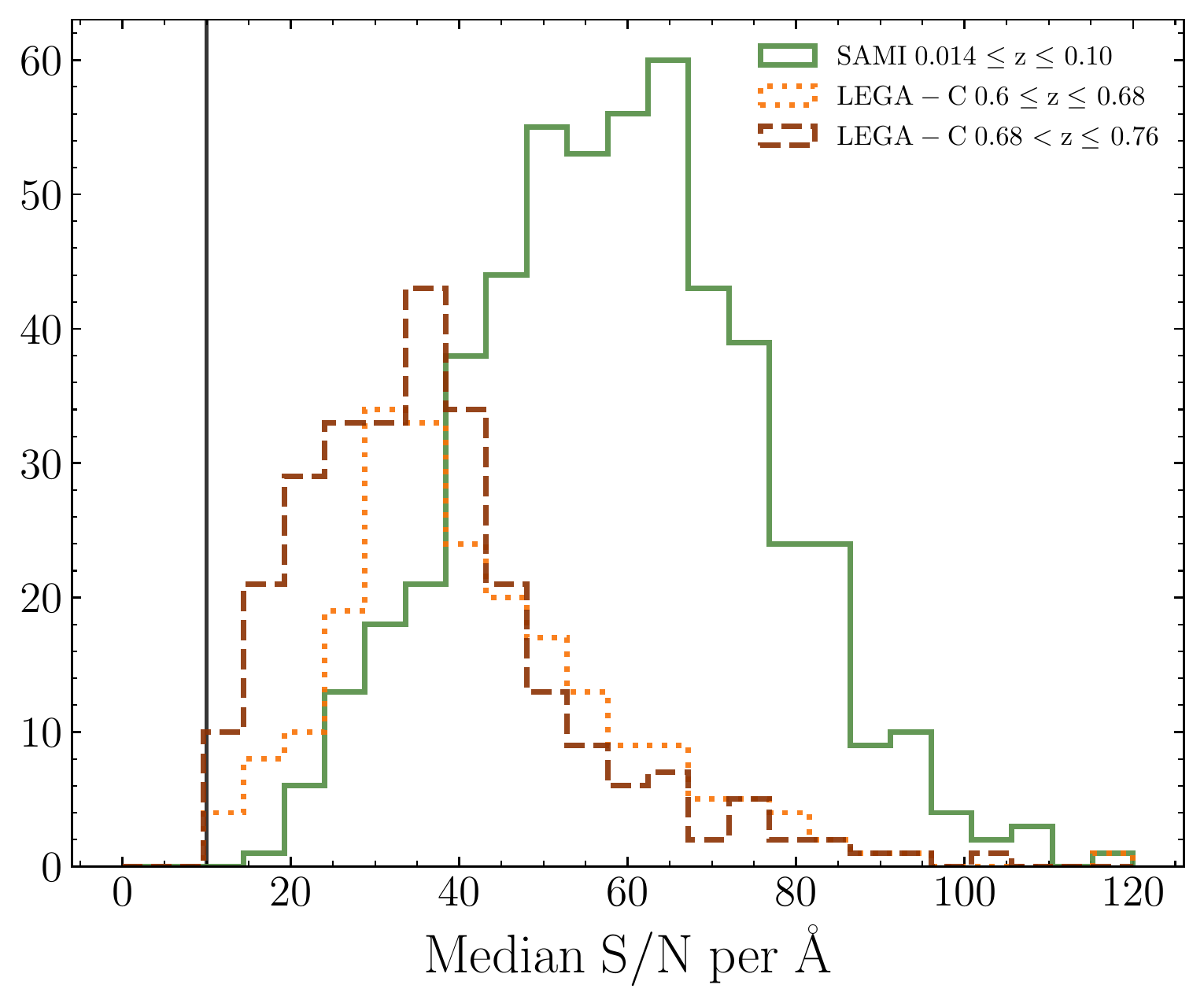}
    \caption[Median spectral signal-to-noise for SAMI and LEGA-C quiescent galaxies]{Median spectral S/N per {\AA} in the rest-wavelength region \wavemin--\wavemax\,{\AA} for SAMI (solid green) and LEGA-C ($z\in[0.60, 0.68]$ dotted orange, $z\in[0.68, 0.76]$ dashed brown) quiescent galaxies. The solid black line represents the S/N threshold (S/N$\geq 10$) of all three redshift samples. The median S/N for the three samples is $\mathrm{S/N_{SAMI}}=59$, and $\mathrm{S/N_{LEGA-C}}=39$ and $35$ for $z\in[0.60, 0.68]$ and $z\in[0.68, 0.76]$ respectively.}
    \label{fig:ks_mag_SN}
\end{figure}

In Figure~\ref{fig:ks_mag_SN} we compare the spectral data quality of the quiescent samples between the two surveys by comparing the median spectral S/N per {\AA} in the rest-wavelength range covered by all targets (\wavemin--\wavemax\,{\AA}). The median S/N for the three samples are $\mathrm{S/N_{SAMI}}=59$, and $\mathrm{S/N_{LEGA-C}}=39$ and $35$ for $z\in[0.60, 0.68]$ and $z\in[0.68, 0.76]$ respectively. The black dotted line shows our adopted quality threshold at S/N $\geq 10$. This minimum S/N requirement only affects 5 quiescent galaxies in the highest redshift LEGA-C bin which span in stellar mass from $\log M_* = 10.6$ to $11.3 M_{\odot}$. We therefore consider the effect of this quality cut on the overall mass-completeness of the sample to be negligible.

\begin{figure*}
    \centering
    \includegraphics[width=\textwidth]{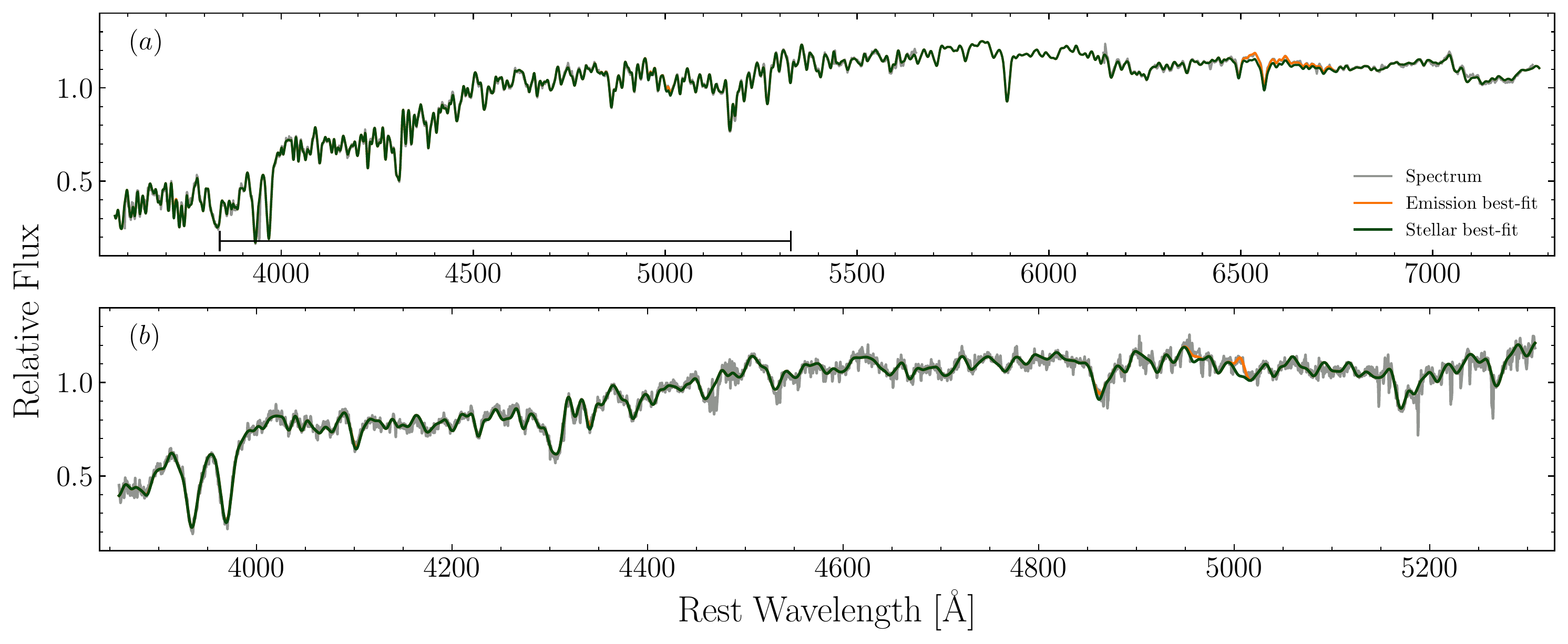}
    \caption[Full spectral fits for SAMI galaxy 278109 and LEGA-C galaxy 206573]{Example full spectral fits for SAMI galaxy ID=278109 (panel~a) and LEGA-C galaxy ID=206573, mask ID=1 (panel~b). In both panels the original spectrum is shown in grey, with the best-fit from the stellar templates in green and the best-fit emission templates in orange. By simultaneously fitting the stellar and gas components, we are able to recover absorption features masked by emission lines. The black horizontal line in panel~a shows the wavelength range of panel~b for ease of comparison. The fluxes are normalised such that the median flux across the spectrum is 1.}
    \label{fig:example_spec}
\end{figure*}

\begin{figure*}
    \centering
    \includegraphics[width=\textwidth]{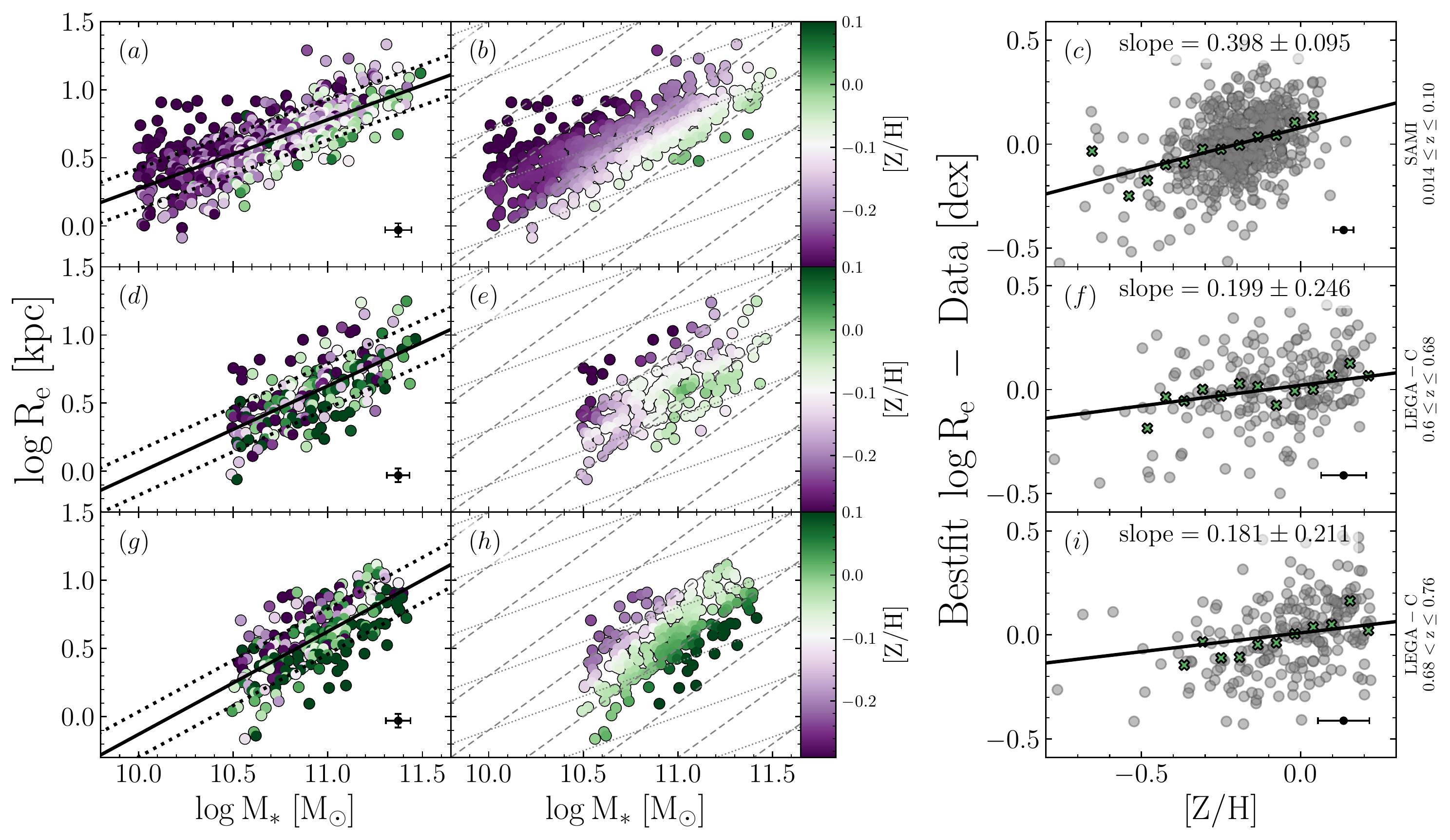}
    \caption[\texorpdfstring{[Z/H]}{} of SAMI and LEGA-C quiescent galaxies in the stellar mass--size plane]{Stellar metallicity [Z/H] of quiescent galaxies in the stellar mass--size plane for three redshift bins. The top row shows the SAMI galaxies $0.014 \leq z \leq 0.10$, middle row the LEGA-C galaxies with $0.60 \leq z \leq 0.68$, and the bottom row the LEGA-C galaxies with $0.68 < z \leq 0.76$. The colour scale of the left column represents metallicity and the centre column shows the smoothed metallicity using the LOESS algorithm. The best-fits are shown in the left column (black solid line) with the fitted $1\sigma$ intrinsic scatter (black dotted lines). The median uncertainty on the mass, size and [Z/H] are shown by the example point in the bottom right corners of the left and right columns. In the centre column, the dashed lines show constant $M_*/R_\mathrm{e}$ and the dotted lines show constant surface density $\Sigma \propto M_*/R_\mathrm{e}^2$. The right column shows the dependence of the residuals (defined as the perpendicular distance between the model minus the data) from the best-fit mass--size relation with metallicity. The black line in the right column is the best-fit relation to the residuals, with the slope written at the top of each panel. The crosses show the median value in independent bins with 3 or more galaxies. This Figure suggests that metallicity varies with $M_*/R_\mathrm{e}$ (dashed lines in the middle column) in all three redshift bins.}
    \label{fig:met_stellar_mass-size_plane}
\end{figure*}

\begin{figure*}
    \centering
    \includegraphics[width=\textwidth]{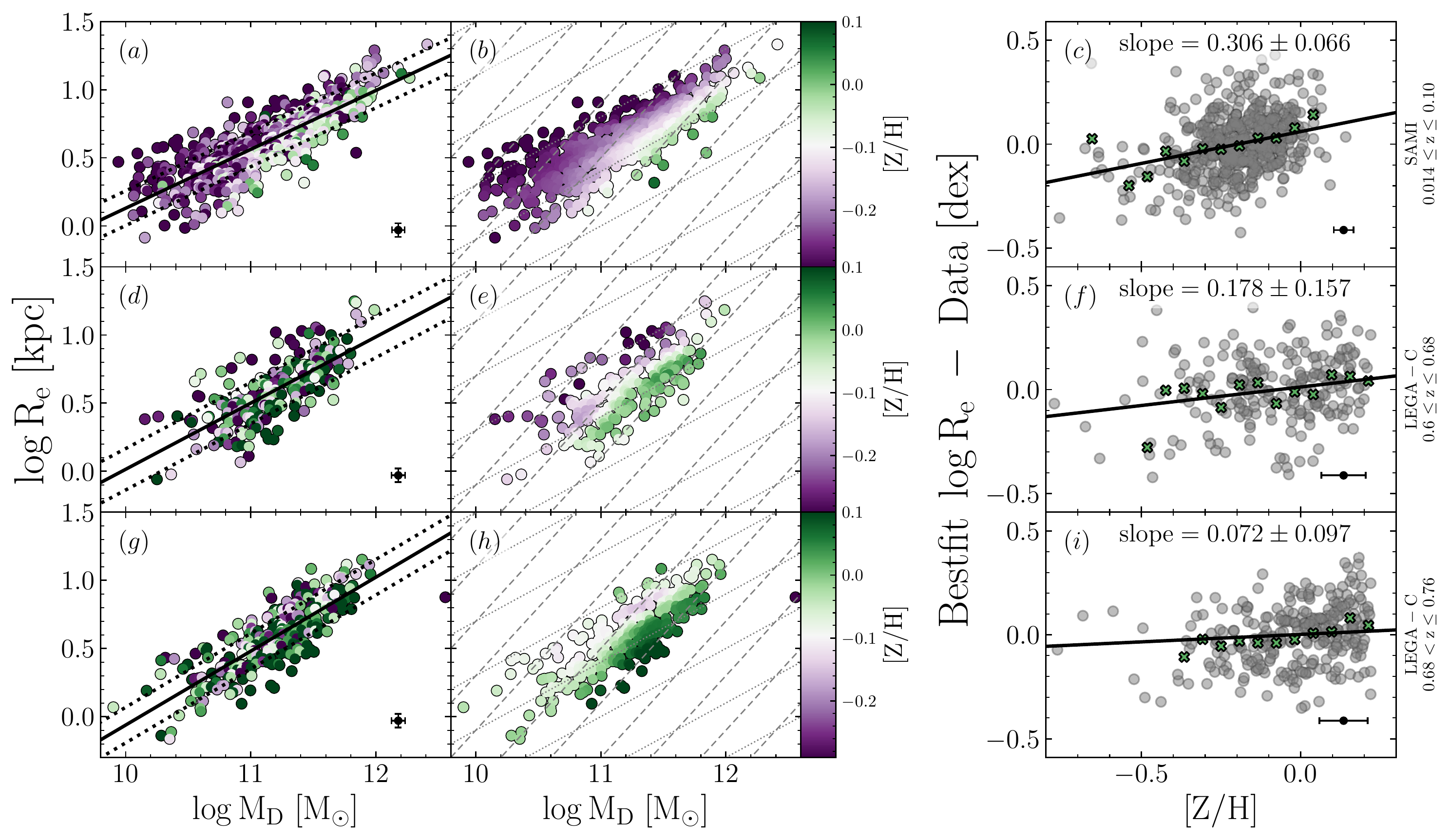}
    \caption[\texorpdfstring{[Z/H]}{} of SAMI and LEGA-C quiescent galaxies in the dynamical mass--size plane]{Stellar metallicity [Z/H] of quiescent galaxies in the dynamical mass--size plane for three redshift bins. The top row shows the SAMI galaxies $0.014 \leq z \leq 0.10$, middle row the LEGA-C galaxies with $0.60 \leq z \leq 0.68$, and the bottom row the LEGA-C galaxies with $0.68 < z \leq 0.76$. The colour scale of the left column represents metallicity and the centre column shows the smoothed metallicity using the LOESS algorithm. The best-fits are shown in the left column (black solid line) with the fitted $1\sigma$ intrinsic scatter (black dotted lines). The median uncertainty on the mass, size and [Z/H] are shown by the example point in the bottom right corner of the left and right columns. In the centre column, the dashed lines show constant $M_D/R_\mathrm{e}$ and the dotted lines show constant surface density $\Sigma \propto M_D/R_\mathrm{e}^2$. The right column shows the dependence of the residuals (defined as the perpendicular distance between the model minus the data) from the best-fit mass--size relation with metallicity. The black line in the right column is the best-fit relation to the residuals, with the slope written at the top of each panel. The crosses show the median value in independent bins with 3 or more galaxies. This Figure suggests that metallicity varies with $M_D/R_\mathrm{e}$ (dashed lines in the middle column) in all three redshift bins.}
    \label{fig:met_dynamical_mass-size_plane}
\end{figure*}

\section{Methodology}\label{sec:methodology}
\subsection{Stellar Population Measurements}\label{sec:stellar_pops_measurement}

Luminosity-weighted stellar population parameters for both the LEGA-C and SAMI galaxies are measured from full spectral fits using 350 theoretical templates based on the Extended Medium resolution INT Library of Empirical Spectra \citep[E-MILES;][]{SanchezBlazquez2006_MILES,Vazdekis2016}, isochrones from \cite{Girardi2000_PADOVA_isochrones} and a \cite{Chabrier2003} initial mass function. We use the full set of 350 templates which contain 50 age values ranging from 0.063 to 17.78 Gyr, 7 [Z/H] values ranging from $-$2.32 to +0.22, and have [$\alpha$/Fe] values scaled to the solar neighbourhood \citep[the "base" models;][]{Vazdekis2016}. The normalisation for the luminosity-weighting is calculated in the range from 3000--8000{\AA}. We tested our analysis using the BaSTI isochrones \citep{Pietrinferni2004,Pietrinferni2006} and found our results qualitatively unchanged. We fit the templates to the de-redshifted spectra using the Python implementation of the publicly available Penalized Pixel-Fitting software \citep[\textsc{pPXF};][]{Cappellari_Emsellem2004,Cappellari2017}. We do an initial fit to ensure that we have a good estimate of the S/N ratio per pixel and multiply the noise by a rescaling coefficient so that the reduced $\chi^2$ is unity. The median rescaling is 1.45 for the LEGA-C sample and 0.54 for the SAMI sample, with standard deviations of 0.31 and 0.21 respectively. We then use this improved estimate of the noise for the second (final) fit. The improved noise estimate is also used to measure the S/N per {\AA} in the rest wavelength range \wavemin--\wavemax\,{\AA}. This wavelength range was chosen because it is covered by all galaxies in the two samples. To account for dust reddening and any offsets to the continuum shape due to flux calibration, this final fit includes a $10^{\rm th}$ degree multiplicative polynomial. 

The final age and metallicity values are the weighted average of all the templates:

\begin{equation}
    \log \mathrm{Age} = \frac{\Sigma w_i \log \mathrm{Age}_i}{\Sigma w_i}
\end{equation}
\begin{equation}
    \mathrm{[Z/H]} = \frac{\Sigma w_i \mathrm{[Z/H]}_i}{\Sigma w_i}
\end{equation}

where $w_i$ is the weight measured from \textsc{pPXF} for the $i$\textsuperscript{th} template, with single-burst age and metallicity values and [Z/H]$_i$ and age$_{i}$. Following \cite{McDermid2015} and \cite{Gonzalez_Delgado2015}, the weighted average age is measured using the logarithm of the template ages because the template ages are sampled  logarithmically.

All fits are done without linear regularisation, which imposes constraints on the weights of neighbouring templates (in age-metallicty space) to vary smoothly. While using linear regularisation produces more realistic star-formation histories, by construction reasonable degrees of regularisation do not significantly change the weighted average age and [Z/H] values \citepalias{Barone2020}.

In addition to the stellar templates, we also include templates for common gas emission lines. We fix the flux ratios of the Balmer series from $H_\alpha$ to $H_\theta$ assuming a decrement based on Case B recombination \citep[electron temperature $T = 10^4\,$K and number density $n = 100\,$cm\textsuperscript{$-3$};][]{Dopita_Sutherland2003}, using the \textsc{tie}\_\textsc{balmer} \textsc{pPXF} keyword. Any residual difference in flux from the fixed ratios is attributed to dust extinction, which is fit using a \cite{Calzetti2000} attenuation curve (via the \textsc{gas\_reddening} keyword). We also limit the ratios of the [OII] and [SII] doublets to lie within the theoretical ranges predicted by atomic physics \citep{Osterbrock_Ferland2006} using the \textsc{limit}\_\textsc{doublets} keyword. The [OIII], [OI] and [NII] doublets also have their flux ratios tied based on atomic physics. We show example fits in Figure~\ref{fig:example_spec} for SAMI galaxy ID=278109 and LEGA-C galaxy ID=206573 mask ID=1. 

The fits are done using the full wavelength range available for each spectrum; the typical rest-wavelength range for the SAMI and LEGA-C spectra are $\sim$3461--7077 {\AA} and $\sim$3723--5168 {\AA} respectively. We tested the resulting age and metallicity values from fits to SAMI spectra that were restricted to use the rest wavelength range typical of the LEGA-C sample. While we found good agreement for both the age and metallicity values, the age values in the reduced wavelength range were on average $0.067 \pm 0.064$~dex younger than when fitted using the full SAMI wavelength range. This modest difference is expected given that the wavelength range 3723--5168~{\AA} contains more and stronger stellar absorption features compared to the range 5168--7077~{\AA}. Given we are interested in the age variations \textit{within} each redshift bin, our results and interpretations are not affected by this minor discrepancy.

We check our stellar population measurements by comparing to other studies probing similar redshift ranges. We note that the absolute age of stellar population templates (and thus resulting age measurements) are not well constrained \citep[e.g.][]{McDermid2015,Scott2017}, therefore, differences in stellar population synthesis methods and models can induce systematic offsets in the absolute age and metallicity value assigned to each galaxy. However the rank order of galaxies relative to each other, and therefore scaling relations, should remain consistent. We find the age--$M_*$ and [Z/H]--$M_*$ relations are consistent in shape across all three redshift bins (with massive galaxies being older and more metal rich), in agreement with the study by \cite{Gallazzi2014} in a similar redshift range. Furthermore, in agreement with \cite{Choi2014}, our measurements show negligible evolution in [Z/H] at fixed mass across the three redshift bins.

We estimate uncertainties for the age and metallicity measurements for both the LEGA-C and SAMI samples by randomly shuffling the residuals between the best-fit and the observed spectrum, which are then reassigned to the best-fit spectrum and refit. This process is repeated 100 times per galaxy, building up a distribution of values in age and metallicity. To ensure we preserve any wavelength dependence of the residuals in this process, the residuals are reassigned within wavelength bins approximately 500\,{\AA} wide. The distributions are approximately Gaussian and centred around the original fit value, so we take the standard deviations of the distributions as the uncertainties on the original age and metallicity measurements. The median uncertainties in the three redshift bins are $\sigma_{\mathrm{[Z/H]}} =$ 0.03, 0.07, and 0.07~dex for metallicity (shown in Figures 5 and 6) and $\sigma_{\log \mathrm{Age}} = $ 0.03, 0.08, and 0.08~dex for $\log \mathrm{Age}$ (shown in figures 9 and 10) for $0.014 \leq z_{\mathrm{SAMI}} \leq 0.10$, $0.60 \leq z_{\mathrm{LEGA-C}} \leq 0.68$, and $0.68 < z_{\mathrm{LEGA-C}} \leq 0.76$ respectively.

For both the SAMI and LEGA-C stellar population fits, we allow for the full range of template ages (0.063 to 17.78 Gyr), despite the age of the Universe at $z=0.76$ being $\sim 7$~Gyr. This is because the absolute age of stellar population templates is not well constrained \citep[e.g.][]{McDermid2015,Scott2017}, and therefore we focus instead on the \textit{relative} difference between age measurements and templates. Using the full set of templates also provides a key way to determine the reliability of our age measurements. 26 (5\%) of LEGA-C age measurements are older than 8 Gyr, indicating that for 95\% of the data the ages are constrained to be younger than the age of the Universe by the data alone, despite allowing for older templates. Additionally, these 26 galaxies have larger age uncertainties than the rest of the sample (mean $\sigma_{\log \mathrm{Age}} = 0.09$~dex for $\log \mathrm{Age} \geq 8$~Gyr, compared to mean $\sigma_{\log \mathrm{Age}} = 0.06$~dex for $\log \mathrm{Age} < 8$~Gyr), therefore these 26 too-old galaxies are already down-weighted in the analysis.

\begin{table*}
\caption{The Spearman correlation coefficient ($\rho$) and $1\sigma$ uncertainty for correlations between stellar population parameter ([Z/H] or Age) and structural parameter $M$, $M/R_\mathrm{e}$ and $M/R^2_\mathrm{e}$ for both the stellar mass ($M_*$) and a dynamical mass  estimate ($M_D \propto \sigma^2 R_\mathrm{e}$). For each set of relations ([Z/H] or Age, $M_*$ or $M_D$) we highlight in bold the structural parameter ($M$, $M/R_\mathrm{e}$, $M/R^2_\mathrm{e}$) with the highest Spearman correlation coefficient (if $\rho > 0.25$ and the difference to the next highest value is greater than 1$\sigma_{\rho}$).}
\begin{tabular}{{cccccccc}}
\hline
Parameter & Sample & $\mathrm{M_*}$ & $\mathrm{M_*/R_e}$ & $\mathrm{M_* /R_e^2}$ & $\mathrm{M_D}$ & $\mathrm{M_D/R_e}$ & $\mathrm{M_D /R_e^2}$ \\
\hline
$\mathrm{[Z/H]}$ & $0.014 \leq \mathrm{z_{SAMI}} \leq 0.10$ & 0.55 $\pm$ 0.01 & $\textbf{0.66 $\pm$ 0.02}$ & 0.36 $\pm$ 0.02 & 0.43 $\pm$ 0.01 & $\textbf{0.55 $\pm$ 0.01}$ & 0.42 $\pm$ 0.02 \\
$\mathrm{[Z/H]}$ & $0.60 \leq \mathrm{z_{LEGA-C}} \leq 0.68$ & 0.26 $\pm$ 0.03 & $\textbf{0.37 $\pm$ 0.04}$ & 0.23 $\pm$ 0.03 & 0.16 $\pm$ 0.03 & $\textbf{0.27 $\pm$ 0.03}$ & 0.21 $\pm$ 0.04 \\
$\mathrm{[Z/H]}$ & $0.68 < \mathrm{z_{LEGA-C}} \leq 0.76$ & 0.27 $\pm$ 0.03 & $\textbf{0.46 $\pm$ 0.03}$ & 0.29 $\pm$ 0.04 & 0.11 $\pm$ 0.03 & 0.30 $\pm$ 0.04 & 0.26 $\pm$ 0.04 \\
Age & $0.014 \leq \mathrm{z_{SAMI}} \leq 0.10$ & 0.08 $\pm$ 0.02 & 0.15 $\pm$ 0.02 & 0.13 $\pm$ 0.02 & 0.21 $\pm$ 0.01 & 0.34 $\pm$ 0.02 & $\textbf{0.40 $\pm$ 0.02}$ \\
Age & $0.60 \leq \mathrm{z_{LEGA-C}} \leq 0.68$ & 0.12 $\pm$ 0.03 & 0.09 $\pm$ 0.04 & 0.01 $\pm$ 0.04 & 0.24 $\pm$ 0.03 & $\textbf{0.30 $\pm$ 0.03}$ & 0.18 $\pm$ 0.04 \\
Age & $0.68 < \mathrm{z_{LEGA-C}} \leq 0.76$ & 0.11 $\pm$ 0.03 & -0.07 $\pm$ 0.04 & -0.11 $\pm$ 0.04 & 0.24 $\pm$ 0.03 & 0.26 $\pm$ 0.04 & 0.05 $\pm$ 0.04 \\
\hline
\end{tabular}
\label{spearman_table}
\end{table*}

\subsection{Linear and Planar fits}
All linear fits are done using a Bayesian approach as outlined by \cite{HoggBovyLang2010} for N points with Gaussian uncertainties on $x$ and $y$ ($\sigma_{x,n}$, $\sigma_{y,n}$), and intrinsic variance in the $y$-direction ($\lambda^2$). Therefore the log-likelihood $\log p$ for slope $m$ and intercept $b$ is:

\begin{equation}\label{equn:2d_likelihood}
    \log p = -0.5 \sum_{n=1}^{N} \left( \frac{\Delta_n^2}{\Sigma_n^2} + \log \Sigma_n^2 \right)
\end{equation}
Where
\begin{align}
    \Delta_n &= y_n - m x_n - b \\
    \Sigma_n &=  (-m, 1) \cdot \begin{pmatrix}
            \sigma_{x,n}^2 & 0 \\
            0 & \sigma_{y,n}^2
            \end{pmatrix} \cdot (-m, 1)^T + \lambda^2
\end{align}
The posterior function takes uniform priors on the angle of the line and the intercept, and a \cite{Jeffreys_prior_1946} prior on the intrinsic scatter in the y-direction ($P (\sigma) \propto 1/\sigma$).

We first find the mode of the posterior function using the differential evolution numerical method \citep{StornPrice1997}. This is followed by Markov Chain Monte Carlo integration of the posterior distribution to estimate the uncertainties on the model parameters, using the Python package \textsc{emcee} \citep{Foreman-Mackey2013}, which implements the affine-invariant ensemble sampler of \cite{GoodmanWeare2010}. A similar Bayesian approach is used for the plane fits in Section~\ref{sec:metallicity_results}, with uniform priors on the slopes and intercepts and a \cite{Jeffreys_prior_1946} prior on the intrinsic scatter in the z-direction (i.e.\ in metallicity). The 2 dimensional likelihood (equation \ref{equn:2d_likelihood}) can be generalised to 3 dimensions with slopes $m_1, m_2$:
\begin{align}
    \Delta_n &= z_n - m_1 y_n - m_2 x_n - b \\
    \Sigma_n &=  (-m_1, -m_2, 1) \cdot \begin{pmatrix}
            \sigma_{x,n}^2 & 0 & 0\\
            0 & \sigma_{y,n}^2 & 0\\
            0 & 0 & \sigma_{z,n}^2
            \end{pmatrix} \cdot (-m_1, m_2, 1)^T + \lambda^2
\end{align}

For illustrative purposes we emphasise the presence/absence of residual trends by smoothing the colour distributions using the LOESS locally weighted regression algorithm \citep{Cleveland_Devlin1988,Cappellari2013b} with a fraction of 0.4 points used in the local approximation. However all fits are done on the unsmoothed data. The LOESS algorithm takes into account the uncertainty on the stellar population parameter (age or metallicity), therefore highly uncertain points may have a significantly different value in the LOESS smoothed colour map compared to the unsmoothed data. We note that these smoothed colour scales are useful to highlight broad age and metallicity variations in the data, but should not be interpreted as reliable for individual datapoints.

\begin{figure}
    \centering
    \includegraphics[width=\columnwidth]{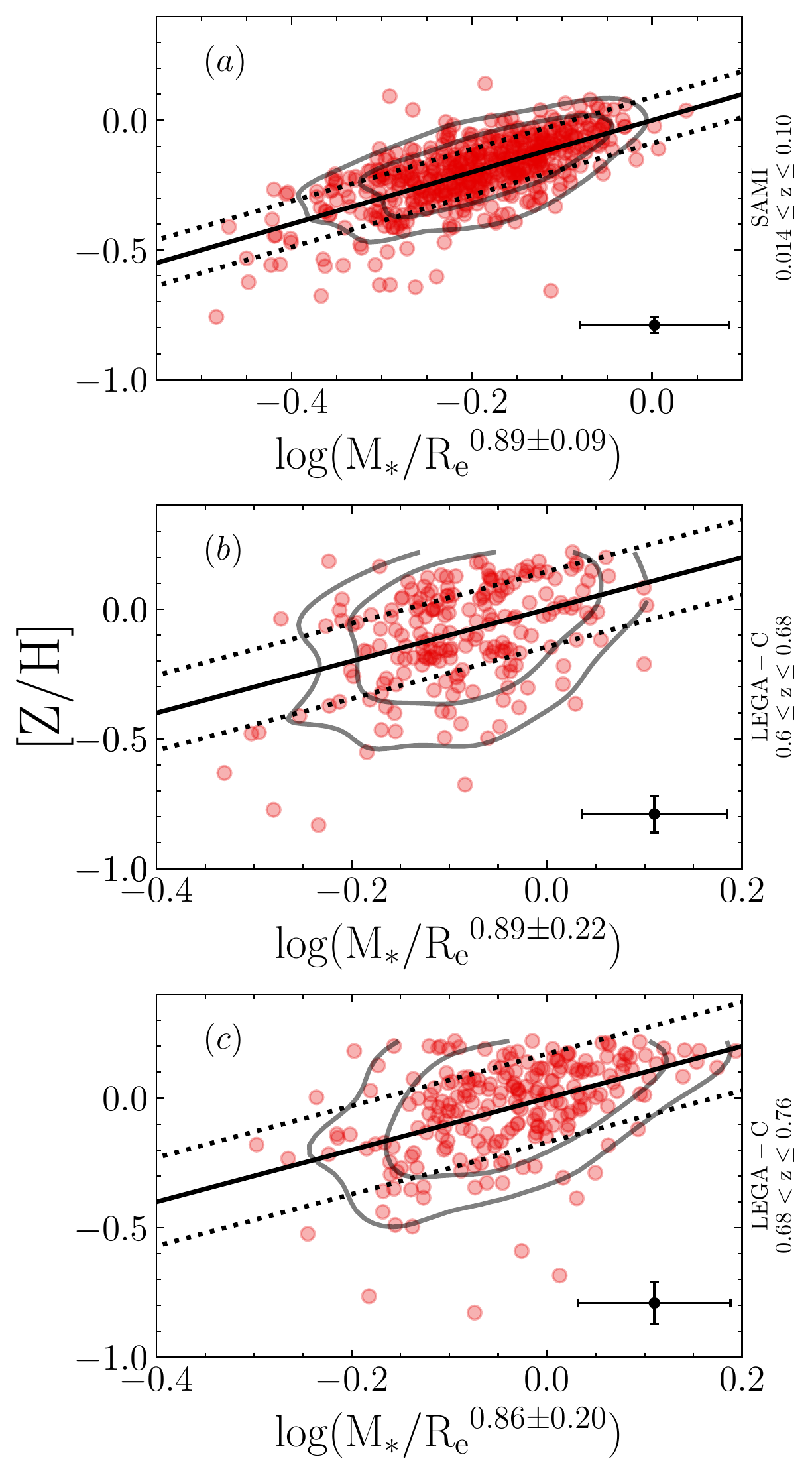}
    \caption[\texorpdfstring{[Z/H]}{} versus best-fit linear combination of $M_*$ and $R_e$]{Stellar metallicity [Z/H] versus the best-fit linear combination of $M_*$ and $R_{\mathrm{e}}$ for the three redshift bins. The top panel (a) shows the SAMI galaxies $0.014 \leq z \leq 0.10$, the middle panel (b) shows the LEGA-C galaxies with $0.60 \leq z \leq 0.68$, and the bottom panel (c) the LEGA-C galaxies with $0.68 < z \leq 0.76$. The best-fits are shown by the black solid line with the fitted $1\sigma$ intrinsic scatter (black dotted lines). The median uncertainty on the $M_*$, $R_\mathrm{e}$ and [Z/H] are shown by the example point in the bottom right corner of each panel. The best-fit ratio of coefficients between $\log M_*$ and $\log R_{\mathrm{e}}$ is statistically consistent within $\sim$2$\sigma$ across all three redshift bins and with a ratio of $-1$.}
    \label{fig:met_stellar_plane_fits}
\end{figure}

\begin{figure}
    \centering
    \includegraphics[width=\columnwidth]{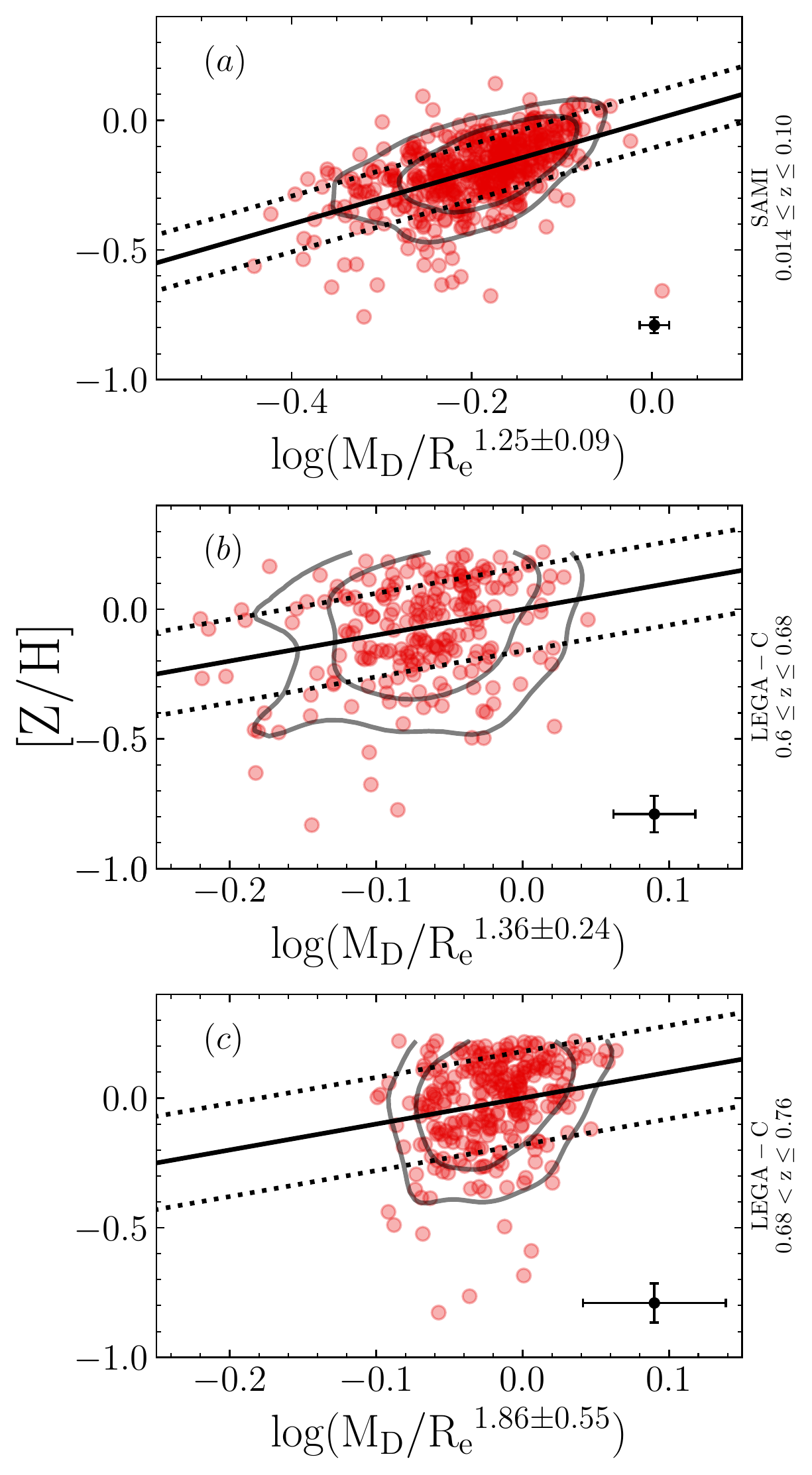}
    \caption[\texorpdfstring{[Z/H]}{} versus best-fit linear combination of $M_D$ and $R_e$]{Stellar metallicity [Z/H] versus the best-fit linear combination of $M_D$ and $R_{\mathrm{e}}$ for the three redshift bins. The top panel (a) shows the SAMI galaxies $0.014 \leq z \leq 0.10$, the middle panel (b) shows the LEGA-C galaxies with $0.60 \leq z \leq 0.68$, and the bottom panel (c) the LEGA-C galaxies with $0.68 < z \leq 0.76$. The best-fits are shown by the black solid line with the fitted $1\sigma$ intrinsic scatter (black dotted lines). The median uncertainty on the $M_D$, $R_\mathrm{e}$ and [Z/H] are shown by the example point in the bottom right corner of each panel. The best-fit ratio of coefficients between $\log M_D$ and $\log R_{\mathrm{e}}$ is statistically consistent within $\sim$3$\sigma$ across all three redshift bins and with a ratio of $-1$.}
    \label{fig:met_dynamical_plane_fits}
\end{figure}

\subsection{Non-Parametric Measure of Correlation}

In Table \ref{spearman_table} we present the Spearman rank-order correlation coefficient along with its $1\sigma$ uncertainty for the relations between stellar population parameters age and [Z/H], and the structural parameters of mass, gravitational potential, and surface-density using both $M_*$ and $M_D$. We estimate the uncertainty on the Spearman coefficients by generating 1000 different datasets consistent with our original measurements. Specifically, each datapoint is sampled 1000 times from a gaussian distribution centered on the datapoint's ``true'' value and with a width equal to the uncertainty on that datapoint. For all relations the distribution of Spearman coefficients is consistent with Gaussian. We therefore use the standard deviation of the distribution as the uncertainty on the coefficient.

\section{Results}\label{sec:results}\label{sec:stellar_pop_results}

\begin{figure*}
    \centering
    \includegraphics[width=\textwidth]{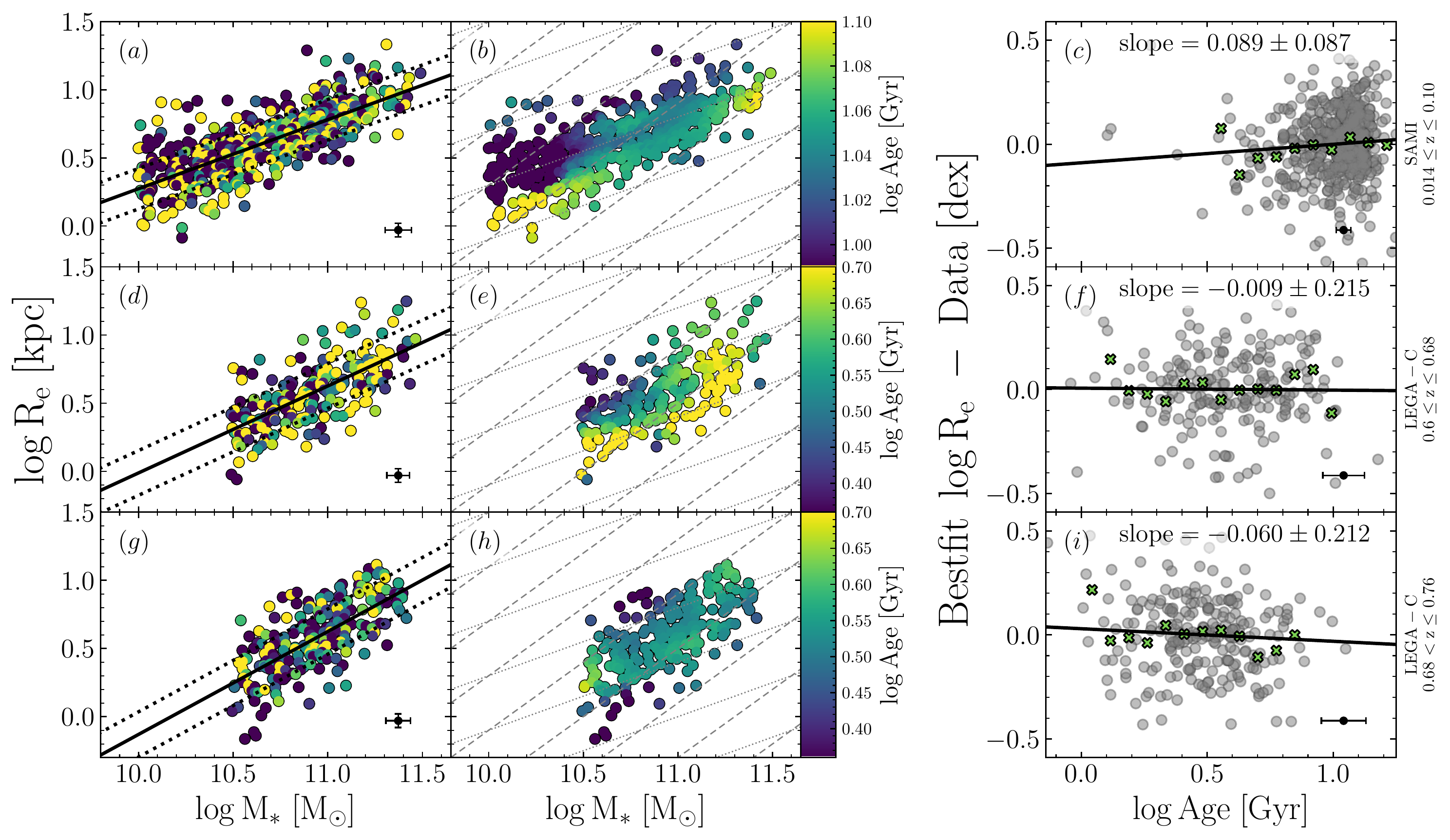}
    \caption[Age of SAMI and LEGA-C quiescent galaxies in the stellar mass--size plane]{Stellar age in the stellar mass--size plane for quiescent galaxies in the three redshift bins. The top row shows the SAMI data with $0.014 \leq z \leq 0.10$, middle row the LEGA-C data with $0.60 \leq z \leq 0.68$, and the bottom row the LEGA-C data with $0.68 < z \leq 0.76$. The colour scale of the left column represents the stellar age, and the centre column shows the smoothed age using the LOESS algorithm. Note that the top row has a different colour scale range then the middle and bottom rows. The best-fits are shown in the left column (black solid line) with the fitted $1\sigma$ intrinsic scatter (black dotted lines). The median uncertainty on the mass, size and age are shown by the example point in the bottom right corner of the left and right columns. In the centre column the dashed lines show constant $M_*/R_\mathrm{e}$ and the dotted lines show constant surface mass density $\Sigma \propto M_*/R_\mathrm{e}^2$. The right column shows the dependence of the residuals (defined as the perpendicular distance between the model minus the data) from the best-fit mass--size relation with age. The black line in the right column is the best-fit relation to the residuals, with the slope written at the top of each panel. The crosses show the median value in independent bins with 3 or more galaxies. This Figure illustrates that variations in age in the mass--size plane change with redshift.}
    \label{fig:age_stellar_mass-size_plane}
\end{figure*}

\begin{figure*}
    \centering
    \includegraphics[width=\textwidth]{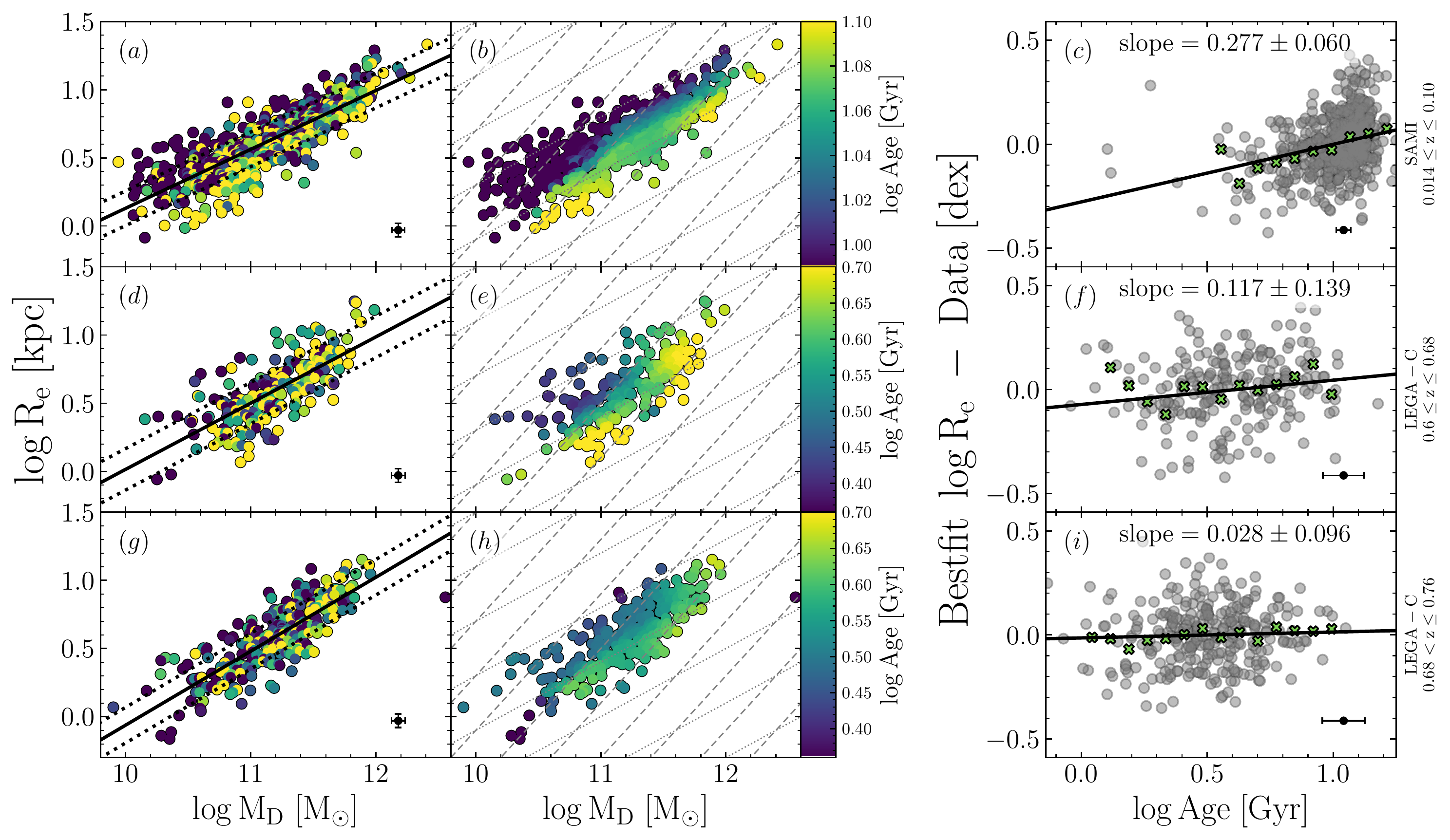}
    \caption[Age of SAMI and LEGA-C quiescent galaxies in the dynamical mass--size plane]{Stellar age in the dynamical mass--size plane for quiescent galaxies in the three redshift bins. The top row shows the SAMI data with $0.014 \leq z \leq 0.10$, middle row the LEGA-C data with $0.60 \leq z \leq 0.68$, and the bottom row the LEGA-C data with $0.68 < z \leq 0.76$. The colour scale of the left column represents the stellar age, and the centre column shows the smoothed age using the LOESS algorithm. Note that the top row has a different colour scale range then the middle and bottom rows. The best-fits are shown in the left column (black solid line) with the fitted $1\sigma$ intrinsic scatter (black dotted lines). The median uncertainty on the mass, size and age are shown by the example point in the bottom right corner of the left and right columns. In the centre column the dashed lines show constant $M_D/R_\mathrm{e}$ and the dotted lines show constant surface mass density $\Sigma \propto M_D/R_\mathrm{e}^2$. The right column shows the dependence of the residuals (defined as the perpendicular distance between the model minus the data) from the best-fit mass--size relation with age. The black line in the right column is the best-fit relation to the residuals, with the slope written at the top of each panel. The crosses show the median value in independent bins with 3 or more galaxies. This Figure illustrates that variations in age in the mass--size plane change with redshift.}
    \label{fig:age_dynamical_mass-size_plane}
\end{figure*}

\subsection{Metallicity [Z/H]}\label{sec:metallicity_results}

We start by showing how luminosity-weighted [Z/H] varies in the mass--size plane using both the stellar mass $M_*$ (Figure~\ref{fig:met_stellar_mass-size_plane}) and the dynamical mass proxy $M_D$ (Figure~\ref{fig:met_dynamical_mass-size_plane}). In all three redshift bins (rows) of Figures~\ref{fig:met_stellar_mass-size_plane} and \ref{fig:met_dynamical_mass-size_plane} we see significant dependence of metallicity on both mass and size. The LOESS-smoothed distributions (middle column) highlight that [Z/H] varies essentially along lines of constant $M/R_\mathrm{e}$ (dashed lines).

The right columns of Figures~\ref{fig:met_stellar_mass-size_plane} and \ref{fig:met_dynamical_mass-size_plane} quantify this variation by showing how the residuals from the best-fit mass--size relation trend with [Z/H], where the residuals are the perpendicular distance between the model minus the data. The low-redshift SAMI sample shows the strongest trend between the mass--size residuals and [Z/H], with panel c in Figures~\ref{fig:met_stellar_mass-size_plane} and  ~\ref{fig:met_dynamical_mass-size_plane} both having $3\sigma$ positive trends (i.e. smaller galaxies at fixed mass are more metal rich). The LEGA-C samples also show positive trends between the residuals and [Z/H], however with decreased statistical significance due to the smaller sample size compared to SAMI.

Table~\ref{spearman_table} further emphasises the dependence of [Z/H] on $M/R_\mathrm{e}$. It shows that, for both $M_*$ and $M_D$, the structural parameter leading to the highest Spearman correlation coefficient is $M/R_\mathrm{e}$. For stellar mass, $M_*/R_\mathrm{e}$ is the highest coefficient by at least $2\sigma_\rho$ for all three redshift samples. Dynamical mass shows a similar result with [Z/H]--$M_D/R_\mathrm{e}$ being highest with over $3\sigma_\rho$ significance for the SAMI sample, although for the LEGA-C samples the [Z/H]--$M_D/R_\mathrm{e}$ coefficient is less than 2$\sigma_\rho$ larger than  [Z/H]--$M_D/R^2_\mathrm{e}$. The [Z/H]--$M_D/R_\mathrm{e}$ coefficients are, however, all 2-3$\sigma_\rho$ larger than [Z/H]--$M_D$, indicating that at fixed mass [Z/H] has a significant dependence on $R_\mathrm{e}$ in all three redshift bins.

We further quantify the dependence of [Z/H] on both mass and size in Figures~\ref{fig:met_stellar_plane_fits} and \ref{fig:met_dynamical_plane_fits}, which show [Z/H] fit as a linear combination of $\log M$ and $\log R_{\mathrm{e}}$, [Z/H]$=a\log M + b\log R_{\mathrm{e}} + c$. In Figure~\ref{fig:met_stellar_plane_fits} the ratio of the coefficients of $\log M_*$ and $\log R_{\mathrm{e}}$ in the best-fit relations are $a/b = -0.89 \pm 0.09$, $-0.89 \pm 0.22$, $-0.86 \pm 0.20$ for $z \in [0.014, 0.10]$, $z \in [0.60, 0.68]$, and $z \in [0.68, 0.76]$ respectively. All three ratios are within 1--2$\sigma$ uncertainty of $-1$, the ratio representing the gravitational potential ($\Phi \propto M_*/R_\mathrm{e}$). Furthermore these ratios are significantly different from 0 (the ratio representing scaling solely with $M_*$) and 2 (scaling with surface density $\Sigma \propto M_*/R^2_\mathrm{e}$).

We find a similar result in Figure~\ref{fig:met_dynamical_plane_fits} using the dynamical mass proxy $M_D$, however we note a slight difference in the optimal coefficient ratio when using $M_*$ and $M_D$; the planar fits with $M_*$ have ratios slightly above $-1$ ($-0.86$ to $-0.89$), while the fits with $M_D$ all have ratios slightly below $-1$ ($-1.25\pm 0.09$, $-1.36 \pm 0.24$, and $-1.86 \pm 0.55$ for $z \in [0.014, 0.10]$, $z \in [0.60, 0.68]$, and $z \in [0.68, 0.76]$ respectively). Although, similarly to Figure~\ref{fig:met_stellar_plane_fits}, all three ratios with $M_D$ are within 1--3$\sigma$ of $-1$ representing the gravitational potential. For the highest redshift bin the ratio $-1.86 \pm 0.55$ is closer to -2, although it is still consistent within $2\sigma$ to -1. Additionally, the Spearman coefficient in the highest redshift bin is slightly higher for [Z/H]--$M_{D}/R_\mathrm{e}$ ($\rho = 0.30 \pm 0.04$) than for $M_{D}/R^2_\mathrm{e}$ ($\rho = 0.26 \pm 0.04$).

\subsection{Age}\label{sec:age_results}

We show how luminosity-weighted age varies in the mass--size plane using both the stellar mass $M_*$ (Figure~\ref{fig:age_stellar_mass-size_plane}) and the dynamical mass proxy $M_D$ (Figure~\ref{fig:age_dynamical_mass-size_plane}). Unlike [Z/H] which shows clear results using either $M_*$ or $M_D$, the age results are clearer when using $M_D$ compared to  $M_*$. We therefore focus on these the results with $M_D$ (Figure~\ref{fig:age_dynamical_mass-size_plane}), and note that the results with $M_*$ are weaker but qualitatively consistent. Overall our results show that stellar age scales with mass and size differently across redshift. For a clear visual representation of the results, we can compare the smoothed colour scale in the middle column from the lowest to the highest redshift bin (middle column, top to bottom row of Figure~\ref{fig:age_dynamical_mass-size_plane}). In the lowest redshift bin there is a significant variation in age perpendicular to the best-fit mass--size relation (panels b), approximately along lines of constant surfance density (dotted lines). This age trend is further highlighted by the right column (panels c) which shows the relation between age and the residuals (orthogonal distance between model $-$ data) from the mass--size best-fit, which shows a positive trend (i.e. smaller galaxies at fixed mass are older). Additionally, Table \ref{spearman_table} shows that the highest Spearman coefficient for the SAMI age relations with $M_D$ is $M_D/R^2_\mathrm{e}$. Figure \ref{fig:age_stellar_mass-size_plane}b shows that age also varies perpendicular to the best-fit stellar mass--size plane at low redshift.

Unlike for the SAMI galaxies, neither LEGA-C redshift bins show a statistically significant trend with the mass--size plane residuals; the slopes in panels (f) and (i) in Figures~\ref{fig:age_stellar_mass-size_plane} and \ref{fig:age_dynamical_mass-size_plane} are all within 1$\sigma$ of zero. From Table \ref{spearman_table} the only LEGA-C age relation with a Spearman coefficient statistically different (at least 1$\sigma_\rho$) from the next highest value is Age--$M_D / R_\mathrm{e}$ for $0.60 \leq z \leq 0.68$. For the highest redshift bin, $0.68 < z \leq 0.76$, the Spearman coefficient for age--$M_D/R_\mathrm{e}$ ($\rho =0.26 \pm 0.04$) and age--$M_D$ ($\rho = 0.24 \pm 0.03$) are consistent within the uncertainties. Therefore, the age goes from trending with surface density ($M/R^2_\mathrm{e}$) at low redshift, to a mild trend with $M/R_\mathrm{e}$ in the intermediate redshift bin, to a weak trend with $M$ or $M/R_\mathrm{e}$ in the highest redshift bin. The change in the correlation between age and $\Sigma$ with redshift suggests the relation observed at $z \sim 0$ is built up over time and that at least one of the processes driving the evolution of mass, size and/or star formation history has a redshift dependence.

Similarly to Figures~\ref{fig:met_stellar_plane_fits} and \ref{fig:met_dynamical_plane_fits} for [Z/H], we performed planar fits to age as a linear combination of $\log M$ and $\log R_\mathrm{e}$. Unlike [Z/H] however, for the LEGA-C age relations, we are unable to disentangle the weak trends between age--$M$ and age--$R_\mathrm{e}$ from the strong $M$--$R_\mathrm{e}$ trend itself. As a result, in a planar fit to $\log\mathrm{Age} = a\log M + b\log R_\mathrm{e} + c$, $\log M$ and $\log R_\mathrm{e}$ are not independent variables. Statistically this is a multicollinearity problem, and the result is that the posterior distributions of the planar fits indicate a wide range of parameter values are plausible, because the fit becomes  $\log \mathrm{Age} = (a + b)\log M + c$ and the coefficients $a$ and $b$ are unconstrained. Therefore we do not include these fits. We note, however, that a similar analysis for a larger subset of SAMI galaxies \citepalias{Barone2018} confirms that, at $z \sim 0$, the most precise predictor of $\log\mathrm{Age}$ is surface mass density (whether photometric or spectroscopic).

We note in particular that in the $z \sim 0$ mass--size plane age trends along lines of constant $\Sigma \propto M/R^2$ (i.e.\ lines with slope = 0.5), which is remarkably similar to the slope of the quiescent population itself in the mass--size plane. This result is evident from the unsmoothed data (panel~a), but clearest in the LOESS-smoothed data in panel~b of Figure~\ref{fig:age_dynamical_mass-size_plane} which shows that the LOESS-smoothed age varies perpendicularly to the mass--size relation. In Section \ref{Population evolution in the Mass--Size Plane} we investigate how the $z \sim 0$ age--$\Sigma$ relation relates to the slope of the quiescent population in the mass--size plane and show how it results from the build-up of the quiescent population over time.

\section{Discussion}\label{sec:discussion}

Our aim was to investigate how scaling relations between galaxy structure and global stellar metallicity and age change across $\sim$6\,Gyr, to understand the redshift dependence in stellar population evolution. We aimed to test two hypotheses: 
\begin{enumerate}
    \item The [Z/H]--$M/R_\mathrm{e}$ relation is consistent with the gravitational potential regulating the retention of stellar and supernova ejecta via its relation to the escape velocity; if this is true, there should be a relation at every redshift.
    \item The $z\sim0$ age--$\Sigma$ relation is built up over time due to galaxies forming and evolving more compactly (diffusely) at higher (lower) redshifts; in this case, the age--$\Sigma$ relation should be less prominent at intermediate redshifts than at $z\sim0$.
\end{enumerate}

Our results support both these hypotheses: (i)~The metallicity of intermediate-redshift quiescent galaxies, like that of low-redshift quiescent \citepalias{Barone2018} and star-forming galaxies \citepalias{Barone2020}, is strongly correlated with $M/R_\mathrm{e}$ (Section~\ref{sec:metallicity_results}); we discuss this further in Section~\ref{sec:Metallicity Discussion}. (ii)~At intermediate redshifts, there is no statistically significant correlation between global stellar age and surface mass density (Section~\ref{sec:age_results}). We show how the $z\sim0$ age--$\Sigma$ is consistent with the redshift-evolution of the mass--size plane in Section~\ref{Population evolution in the Mass--Size Plane} and further discuss this hypothesis in Section~\ref{sec:Age Discussion}.

\subsection{The consistency of the [Z/H]--\texorpdfstring{$\Phi$}{} relation across 6\texorpdfstring{\,}{}Gyr}\label{sec:Metallicity Discussion}

We find that from low to intermediate redshift ($z \leq 0.76$) the stellar metallicity of quiescent galaxies correlates more tightly with $M/R_\mathrm{e}$ than with other combinations of mass and size. This result is in agreement with the low-redshift results of \citetalias{Barone2018} and \citetalias{Barone2020} for the stellar metallicity of quiescent and star-forming galaxies respectively and \cite{DEugenio2018} for the gas-phase metallicity of star-forming galaxies. Our results also agree with \cite{Diaz-Garcia2019}, who showed that (at fixed mass) smaller quiescent galaxies are more metal-rich since $z\sim 1$. We note that, for non-zero uncertainty on $R_{\mathrm{e}}$, $M_*/R_{\mathrm{e}}^x$ must have a higher observational uncertainty than $M_*$ alone (for $x \neq 0$). Therefore, our result that the observed [Z/H]--$M_*/R^{\sim1}$ relation is tighter than the relation with $M_*$ alone means the relation with $M_*/R^{\sim1}$ must be \textit{intrinsically} tighter.

These studies (\citetalias{Barone2018}, \citealt{DEugenio2018}, \citetalias{Barone2020}) built on earlier works \citep[e.g.][]{Worthey_etal1992,Trager2000b} that proposed the global stellar and gas-phase metallicity of a galaxy is regulated by the gravitational potential ($\Phi$), as the depth of the potential well determines the escape velocity required for metal-rich gas to be expelled from the system by supernova, active galactic nuclei (AGN) and/or stellar winds and thus avoid being recycled into later stellar generations. Our finding that the relation also exists at $0.60 \leq z \leq 0.76$ supports this hypothesis.

It is interesting to note that the ratio of coefficients of the best-fit planes for the three samples are all consistent within 1$\sigma$ (for either $M_*$ or $M_D$). If we simplify the complex process of metal production and outflows to a closed box model and assume the global stellar metallicity results from a balance between metal production determined by the stellar mass and metal retention determined by the gravitational potential and the strength and frequency of gas outflows, our results are consistent with there being no significant change in the balance of these processes from $0.014 \leq z \leq 0.76$. While we are unable to quantitatively test this hypothesis due to differences in the sample selection between the two surveys, future works focusing on the slope and scatter of the [Z/H]--$\Phi$ relation across redshift may shed light on this topic, and help place constraints on the strength and efficiency of feedback process in heating and expelling interstellar gas.

\begin{figure*}
    \centering
    \includegraphics[width=\textwidth]{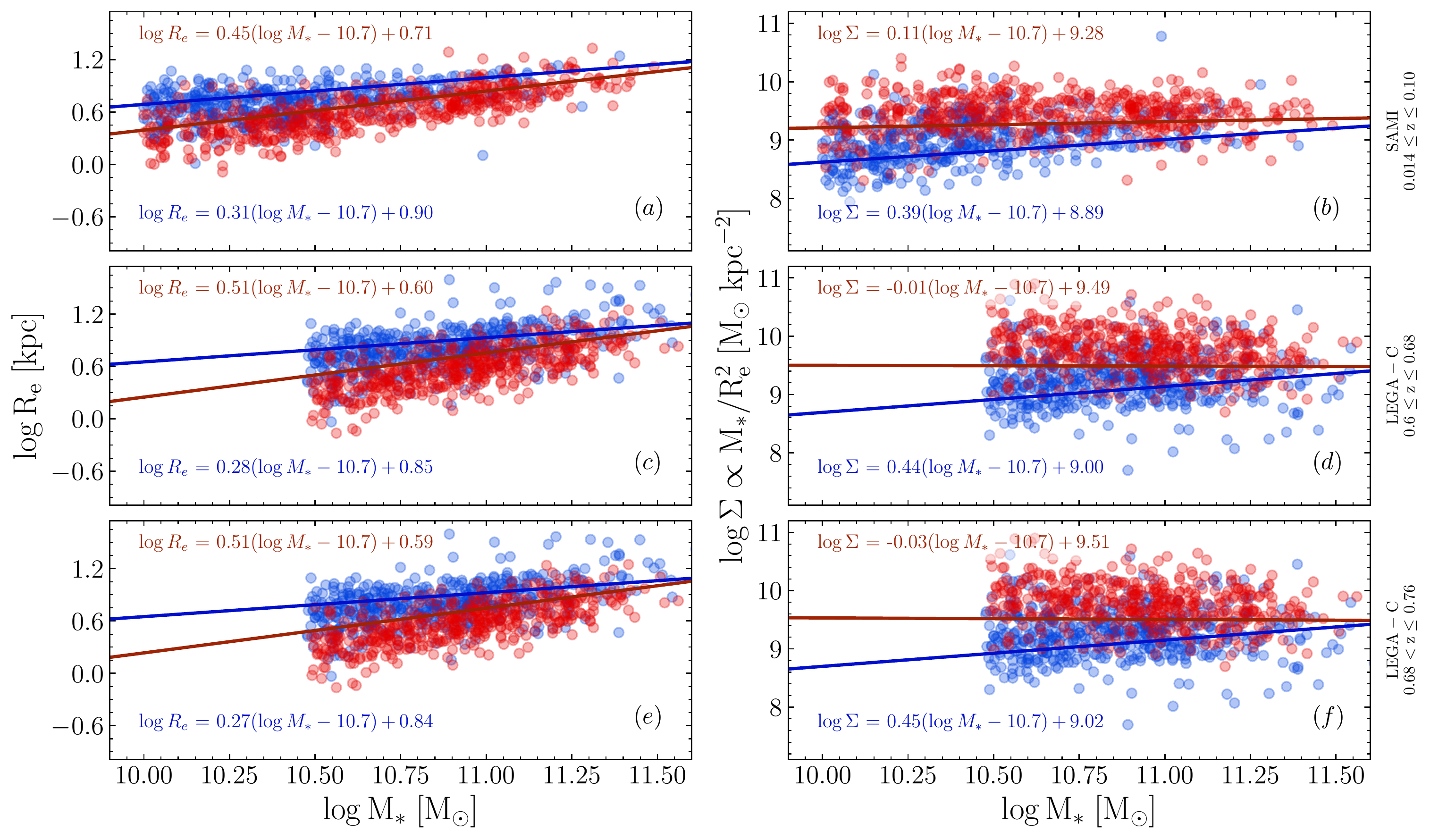}
    \caption[Mass--size and mass--surface density relations for SAMI and LEGA-C galaxies]{Mass--size (left column) and mass--surface density (right column) planes for the three redshifts bins. The top row shows the SAMI data with $0.014 \leq z \leq 0.10$, middle row the LEGA-C data with $0.60 \leq z \leq 0.68$, and the bottom row the LEGA-C data with $0.68< z\leq 0.76$. The red points are quiescent galaxies and the blue points are star-forming galaxies. In the left columm, the solid red and blue lines are the redshift-dependent mass--size relations to the quiescent and star-forming populations based on the analysis of \cite{Mowla2019}. The equations for these relations are written in each panel (in red at the top for quiescent galaxies and in blue at bottom for star-forming galaxies). In the right column, we show the same mass--size relations from \cite{Mowla2019} converted to the mass--surface density plane.}
    \label{fig:mass-size_mass-surfacedensity_planes}
\end{figure*}

\subsection{The build-up of the age--\texorpdfstring{$\mathbf{\Sigma}$}{} relation over 6\texorpdfstring{\,}{}Gyr}\label{sec:Age Discussion}

Unlike the metallicity results, we find a significant difference between the age distribution in the mass--size plane at intermediate redshift compared to low redshift. \citetalias{Barone2018} and \citetalias{Barone2020} found that the global age of both low-redshift early-type and star-forming galaxies tightly correlates with the surface density ($\Sigma \propto M/R_{\mathrm{e}}^2$), whereas we find no significant correlation with $\Sigma$ in either LEGA-C redshift sample. Comparing the results from the highest to lowest redshift bin, age changes from varying weakly with $M$ and/or $M/R_\mathrm{e}$ ($0.68 < z \leq 0.76$), to weakly correlating with $M/R_\mathrm{e}$ ($0.6 \leq z \leq 0.68$), to a significant variation with $M/R^2_\mathrm{e}$, approximately \textit{perpendicular} to the mass--size relation ($0.014 \leq z \leq 0.10$). This suggests that, unlike the [Z/H]--$\Phi$ relation, the $z \sim 0$ age--$\Sigma$ does not reflect a causal relation, but is instead built up over time due to the fact that both age and $\Sigma$ depend on the conditions of the Universe when a galaxy becomes quiescent. This hypothesis is supported by a number of studies linking age, $\Sigma$, and formation epoch. In particular, \cite{Kauffmann2003_II,Kauffmann2006} and \cite{Franx2008} showed that galaxy star formation histories are strongly correlated with surface mass density. This correlation between star formation history and surface density was further refined by \cite{Diaz-Garcia2019} and \cite{Zolotov2015}, who showed that central surface density correlates strongly with formation epoch and quenching epoch respectively. We build upon these previous work to relate the change across redshift of the age dependence on mass and size to the evolution of the mass--size relation in the following Section \ref{Population evolution in the Mass--Size Plane}, and show that we can explain the $z \sim 0$ age--$\Sigma$ relation as being due to the average surface density of galaxies in both the star-forming and quiescent populations decreasing with decreasing redshift.

\subsubsection{Age in the mass--size and mass--surface density planes}\label{Population evolution in the Mass--Size Plane}

It is well established that the median size of galaxies in both the quiescent and star-forming populations evolves significantly in the mass--size plane with redshift \citep{Ferguson2004,Trujillo2007,Buitrago2008,Williams2010}. Specifically, \cite{vanderWel2014} and \cite{Mowla2019} showed that for $0 \leq z \leq 3$, the slopes of the mass--size relations for both the quiescent and star-forming populations change little with redshift, but the intercepts decrease with increasing redshift such that high-redshift populations are smaller at fixed mass. 

We illustrate this result in the left column of Figure~\ref{fig:mass-size_mass-surfacedensity_planes}, which shows our quiescent and star-forming populations in the mass--size plane for each redshift bin. The solid blue and red lines are the redshift-dependent best-fit stellar mass--size relations from \cite{Mowla2019}. The left column of Figure~\ref{fig:mass-size_mass-surfacedensity_planes} shows that, while our quiescent galaxy mass estimates are systematically slightly below the \cite{Mowla2019} relations, our samples are in overall good agreement.

Of crucial importance are the slopes of mass--size relations for the quiescent populations. In all three redshift bins the mass--size relation for the quiescent population is approximately $\log R_{\mathrm{e}} \approx 0.5 \log M_*$ (specifically, $0.45$, $0.51$ and $0.51$ for $z \in [0.014, 0.10]$, $z \in [0.60, 0.68]$ and $z \in [0.68, 0.76]$ respectively). This slope of $\sim$0.5 in the mass--size plane means that, at fixed redshift, the quiescent population has an approximately constant surface--mass density regardless of stellar mass. For some redshift-dependent constant $k_1(z)$,
\begin{align}
    \log R_{\mathrm{e}} &= 0.5 \log M_* + k_1(z) \\
    \log \Sigma &= \log M_* - 2\log R_{\mathrm{e}} = -2k_1(z)
\end{align}
which implies that surface density does not change with stellar mass:
\begin{align}
    \frac{\delta \log \Sigma}{\delta \log M_*} &= 0
\end{align}

Indeed, in the right column of Figure~\ref{fig:mass-size_mass-surfacedensity_planes} we show our samples in the stellar mass--surface density plane. The red and blue lines are the mass--size relations from \cite{Mowla2019} converted to be in terms of mass and surface density. As expected given the mass--size relations in the left column, the slopes of the quiescent populations are all close to zero (0.11, -0.01, and -0.01 for $z \in [0.014, 0.10]$, $z \in [0.60, 0.68]$ and $z \in [0.68, 0.76]$). Furthermore, as expected, the intercepts in the mass--surface density plane increase with increasing redshift, such that the LEGA-C quiescent galaxies are more compact (higher $\Sigma$) than the lower redshift SAMI quiescent galaxies.

The individual evolutionary tracks of star-forming galaxies are expected to closely align with the star-forming population in the mass--size plane, a result supported by both observations and simulations \citep{Lilly1998,Ravindranath2004,Trujillo2006_apj,vanDokkum2015}. Therefore the slope of $\sim$0.3 for the star-forming sequence in the mass--size plane in all three redshift bins means that as galaxies build up their mass via in-situ star formation, they also increase their surface density. For some redshift-dependent constant $k_2(z)$, 
\begin{align}
    \log R_\mathrm{e} &= 0.3 \log M_* + k_2(z) \\
    \log \Sigma &= \log M_* - 2\log R_\mathrm{e} = 0.4 \log M_* - 2 k_2(z)
\end{align}
which implies surface density increases with increasing stellar mass:  
\begin{align}
    \frac{\delta \log \Sigma}{\delta \log M_*} &= 0.4
\end{align}

Combining these two key results suggests that star-forming galaxies build their stellar mass, and at the same time increase their stellar surface density until they reach a redshift-dependent threshold surface density at which they quench. As this threshold surface density decreases with decreasing redshift, the quiescent population is built up to include galaxies that have quenched at a range of redshifts and, consequently, with a range of surface densities. In the $z \sim 0$ quiescent population, therefore, a galaxy's surface density reflects the redshift at which it quenched (luminosity-weighted age is closely tied to how long ago the galaxy quenched). 

This connection between quenching and galaxy structure, in particular surface density, is in agreement with previous work. \cite{Franx2008} found a redshift-dependent threshold surface density above which galaxies have low sSFR (are quiescent) and below which the sSFR are high with little variation (are star-forming). \cite{Franx2008} also found that the surface density threshold increases with increasing redshift. Similarly, \cite{vanderWel2009} found that stellar velocity dispersion also shows a redshift-dependent threshold separating quiescent and star-forming galaxies. Additionally, \cite{Gonzalez_Delgado2014_I} found a threshold surface-mass density for low-redshift spheroidal galaxies that is nearly independent of stellar mass, in agreement with our results.

Recent work by \cite{Chen2020} provides a theoretical framework explaining the relation between surface density and quenching. \cite{Chen2020} showed that a galaxy model in which central black hole mass (and therefore strength of AGN feedback) is related to both the host galaxy's mass and size successfully explains key properties of star-forming and quenched galaxies. In the model larger star-forming galaxies at fixed mass have smaller central super-massive black holes (due to their lower central surface density). Therefore these extended galaxies evolve to higher stellar masses before the central AGN has strong enough feedback to shock-heat infalling gas from the halo, quenching star formation.

There remain two unanswered questions: (1)~what leads to galaxies forming more compactly at higher redshifts? and (2)~what leads to the decrease with time of the redshift-dependent threshold surface density at which star-forming galaxies quench? In relation to the first question, \cite{Franx2008} concluded that compact galaxies must have formed their stars earlier when the Universe was denser and had a higher gas fraction compared to the formation epoch of galaxies with lower surface densities. \cite{Wellons2015} used the Illustris simulations \citep{Genel2014,Vogelsberger2014} to trace the star-formation and assembly histories of compact galaxies and found that another dominant mechanism for forming compact galaxies at high redshift are centrally-concentrated starbursts triggered by wet major mergers. \cite{Wellons2015} noted that this wet major merger mechanism is also intrinsically linked to the density and redshift evolution of the Universe; as the Universe expanded and the abundance of cold gas decreased, the likelihood of wet major mergers also decreased with decreasing redshift. While these explanations may underlie the change in the formation and evolution of galaxies across redshift (question~1), they do not explain the origin of the redshift-dependent surface density threshold at which star-forming galaxies quench (question~2). This changing threshold may be due a redshift-dependence of the complex balance between the supply and heating of inflowing gas and the feedback and outflows from star formation and AGN \citep[e.g.,][]{Chen2020}.

Clearly, this scenario of a redshift-dependent quenching surface-density requires that passive evolution does not significantly alter the inner structure of quiescent galaxies (inside $\sim$1~$R_\mathrm{e}$); however, we note that if this assumption was incorrect, it would be even harder to explain the observed correlation between stellar population age and surface mass density.

\subsubsection{The effect of mergers}\label{mergers_effect}

Thus far we have only considered the evolution in the average properties of the quiescent and star-forming populations due to the addition (and loss) of new (and old) members. However individual galaxies also grow in mass and size through both minor \citep[mass ratio $\lesssim 0.3$ e.g.][]{Lambas2012} and major mergers \citep{Oser2012,Bluck2012,Oogi_Habe2013,Ownsworth2014}. Indeed, the relative absence at low redshift of the very compact galaxies frequently observed at high redshift indicates individual galaxies must undergo significant mass--size evolution even after ceasing star formation \citep{vanDokkum2010}. Dry major mergers are expected to increase a galaxy's mass and size proportionally, whereas minor mergers significantly increase a galaxy's effective radius while contributing comparatively little to its stellar mass \citep{Bezanson2009,Naab2009,Hopkins2009}. Given these two relations, both major and minor mergers will increase a galaxy's mass but \textit{decrease} its surface mass density and therefore move galaxies down and right in the mass--$\Sigma$ plane. Specifically, based on results from \cite{Naab2009} and \cite{Bezanson2009}:
\begin{align}
    \Delta \log R_{\mathrm{e},\rm major} &\approx \Delta \log M_* \\
    \Delta \log R_{\mathrm{e},\rm minor} &\approx 2\Delta \log M_*
\end{align}
Therefore the changes in surface mass density are always negative (galaxies become more diffuse):
\begin{align}
    \Delta \log \Sigma_{\rm major} &\approx - \Delta \log M_* \\
    \Delta \log \Sigma_{\rm minor} &\approx - 3\Delta \log M_*
\end{align}

We can explore the implications of mergers on the build-up of the age--surface density relation by considering a hypothetical scenario in which galaxies quench at a surface density \textit{independent} of redshift (i.e.\ there is no redshift evolution of the mass--size plane). If we consider only mass growth through mergers (i.e.\ no star-formation) and assume that older galaxies (those that formed earlier) will on average have undergone more mergers than younger galaxies, this would lead to older galaxies having on average a \textit{lower} surface density than young galaxies at fixed mass---the opposite of the trend seen at $z \sim 0$. Therefore, while individual galaxy evolution through mergers will undoubtedly introduce scatter into the relation, it cannot be the cause of our age results; in fact, it suggests the original relation must be even stronger than observed today. In addition to increasing a galaxy's mass and size, mergers also influence global stellar population properties. In particular the old, metal rich stellar populations of massive compact galaxies are diluted with younger metal-poor stars when merging with smaller, more diffuse systems. Furthermore, mergers can also restart star-formation in quenched galaxies, lowering the mean stellar population age and metallicity. At $z<1$ between 10--15\% of quiescent galaxies are estimated to have undergone rejuvenated star formation \citep{Thomas2010,Chauke2019}, which may account for some of the intrinsic scatter in the age--$\Sigma$ correlation.

\section{Summary and Conclusions}\label{sec:Summary and Conclusions}

In this study we aimed to test two key hypotheses on the origins of low redshift stellar population scaling relations:

\begin{enumerate}
    \item The [Z/H]--$M/R_\mathrm{e}$ relation at $z\sim 0$ is due to the gravitational potential regulating the retention of stellar and supernova ejecta via its relation to the escape velocity. If so, there should also be a tight correlation between [Z/H] and $M/R_\mathrm{e}$ at intermediate ($z\sim 0.7$) redshifts.
    \item The age--$\Sigma$ relation at $z\sim0$ is built up over time due to galaxies forming and evolving more compactly (diffusely) at higher (lower) redshifts. If true, at intermediate redshifts the age--$\Sigma$ relation should be less prominent.
\end{enumerate}

To achieve these goals, we used \textsc{pPXF} and the E-MILES library of synthetic stellar templates to measure global light-weighted stellar ages and metallicities for a representative sample of quiescent galaxies spanning 6~Gyr of cosmic history. The data consists of 524 quiescent galaxies with $0.014 \leq z \leq 0.10$ from the SAMI Galaxy Survey, and 492 quiescent galaxy split between $0.60 \leq z \leq 0.68$ and $0.68 < z \leq 0.76$ from the LEGA-C Survey. We quantified how the global stellar population parameters of age and metallicity vary in the mass--size plane between low ($0.014 \leq z \leq 0.10$) and intermediate ($0.60 \leq z \leq 0.76$) redshifts. Specifically, we investigated whether the [Z/H]--$M/R_\mathrm{e}$ and age--$M/R_\mathrm{e}^2$ relations found at low redshift by \citetalias{Barone2018} for quiescent galaxies are also present in a sample with similar masses at a lookback time of 6\,Gyr. We find the [Z/H]--$M/R_\mathrm{e}$ relation is also present at intermediate redshifts but the age--$M/R_\mathrm{e}^2$ relation is not, in agreement with both our hypotheses.

Our conclusion that the [Z/H]--$M/R_\mathrm{e}$ also exists at $0.60 \leq z \leq 0.76$ extends our previous results at low redshift for stars \citepalias{Barone2018,Barone2020} and gas \citep{DEugenio2018} and supports the theory that the depth of the gravitational potential well regulates the stellar and gas-phase metallicity by determining the escape velocity required for metal-rich gas to be expelled from the system and thus avoid being recycled into later stellar generations \citep[e.g.][]{Franx1990}.

To understand the change in the way age varies across the mass--size plane from low to intermediate redshift, we consider the evolution of the mass--size plane itself. Specifically, we show that the slope of $\sim$0.5 for the quiescent population in the mass--size plane in this redshift range \citep{Mowla2019} leads to flat slopes in the mass--surface density plane, with an intercept that decreases with decreasing redshift. This implies that star-forming galaxies reach a redshift-dependent threshold surface density at which they quench. Importantly this threshold is higher at higher redshifts, so that galaxies forming and evolving at higher redshifts reach a higher surface density before quenching compared to low-redshift star-forming galaxies. The age--surface density relation at $z \sim 0$ is therefore the result of the build-up of the low redshift quiescent and star-forming populations from galaxies that have formed, evolved, and quenched over a range of redshifts, and hence over a range of surface densities. Consequently, the age--surface density relation at $z\sim0$ arises from the cumulative effect of the redshift-dependent processes that drive the evolution of the star-forming and quiescent populations in the mass--size plane.

Future spectroscopic surveys such as MOONRISE \citep{Maiolino2020_MOONRISE} will help to push this relation to higher redshift ($z\sim 1 - 2.5$) while the large sample size and mass range of the Hector survey \citep{Bryant2020_Hector} at $z\sim 0$ will allow exploration of these relations in the low-mass regime.

\section{Acknowledgements}
We thank the anonymous referee for their attention to detail which helped improve the paper. TMB is supported by an Australian Government Research Training Program Scholarship. FDE and AvdW acknowledge funding through from the European Research Council (ERC) under the European Union’s Horizon 2020 research and innovation program, grant agreement No. 683184. FDE also acknowledges funding through the ERC Advanced grant 695671 "QUENCH" and support by the Science and Technology Facilities Council (STFC). NS acknowledges the support of an Australian Research Council Discovery Early Career Research Award (project number DE190100375). JvdS acknowledges support of an Australian Research Council Discovery Early Career Research Award (project number DE200100461) funded by the Australian Government. P.F.W. acknowledges the support of the fellowship from the East Asian Core Observatories Association. RB gratefully acknowledges funding provided by the Robert C. Smith Fund and the Betsy R. Clark Fund of The Pittsburgh Foundation. LC is the recipient of an Australian Research Council Future Fellowship (FT180100066) funded by the Australian Government. JBH is supported by an ARC Laureate Fellowship FL140100278. The SAMI instrument was funded by Bland-Hawthorn's former Federation Fellowship FF0776384, an ARC LIEF grant LE130100198 (PI Bland-Hawthorn) and funding from the Anglo-Australian Observatory. JJB acknowledges support of an Australian Research Council Future Fellowship (FT180100231). MSO acknowledges the funding support from the Australian Research Council through a Future Fellowship (FT140100255).

This research was supported by the Australian Research Council Centre of Excellence for All Sky Astrophysics in 3 Dimensions (ASTRO 3D) through project CE170100013.

The LEGA-C Public Spectroscopy Survey observations were made with ESO Telescopes at the La Silla Paranal Observatory under program IDs 194-A.2005 and 1100.A-0949.

The SAMI Galaxy Survey is based on observations made at the Anglo-Australian Telescope. The Sydney-AAO Multi-object Integral field spectrograph (SAMI) was developed jointly by the University of Sydney and the Australian Astronomical Observatory. The SAMI input catalogue is based on data taken from the Sloan Digital Sky Survey, the GAMA Survey and the VST ATLAS Survey. The SAMI Galaxy Survey is supported by the Australian Research Council Centre of Excellence for All Sky Astrophysics in 3 Dimensions (ASTRO 3D) through project number CE170100013, the Australian Research Council Centre of Excellence for All-sky Astrophysics (CAASTRO), through project number CE110001020, and other participating institutions. The SAMI Galaxy Survey website is http://sami-survey.org/.

We acknowledge the traditional custodians of the land on which the AAT stands, the Gamilaraay people, and pay our respects to their elders past and present.

This research was conducted using the freely available \textsc{Python} programming language \citep{vanRossum1995}, maintained and distributed by the Python Software Foundation, and the \textsc{IPython} extension \citep{IPython}. Our analysis made use of the \textsc{NumPy} \citep{NumPy}, \textsc{SciPy} \citep{scipy}, \textsc{Astropy} \citep{Astropy}, \textsc{matplotlib} \citep{Hunter2007}, and \textsc{corner} \citep{Foreman-Mackey_cornerplot2016} packages. In preliminary analyses, we also used \textsc{TOPCAT} \citep{Taylor2005} and \textsc{Ned Wright's Cosmology Calculator} \citep{Wright2006}. We make extensive use of NASA’s Astrophysics Data System.

\section*{Data Availability}
The SAMI data used in this work is publically available and hosted on the Australian Astronomical Optics’ Data Central \citep[https://datacentral.org.au/][]{Croom2021_SAMIDR3}. The LEGA-C data used is in the public domain via the ESO Science Archive and through the Mikulski Archive for Space Telescopes (MAST). The full reduced data is available in the third public data release. Details of the data release are hosted on the ESO website \citep[http://archive.eso.org/cms/eso-archive-news/Third-and-final-release-of-the-Large-Early-Galaxy-Census-LEGA-C-Spectroscopic-Public-Survey-published.html][]{vanderWel2021}.

\bibliographystyle{mnras}

\bsp 
\label{lastpage}
\end{document}